\pdfoutput=1
\documentclass[aps,pra,twocolumn,groupedaddress,showpacs]{revtex4}
\usepackage{graphicx} 
\usepackage{amsmath} 
\begin{document}


\title{Observation of St\"{u}ckelberg oscillations in dipole-dipole interactions}


\author{C. S. E. van Ditzhuijzen}
\author{Atreju Tauschinsky}
\author{H. B. van Linden van den Heuvell}
\affiliation{Van der Waals-Zeeman Institute, University of Amsterdam, Valckenierstraat 65, 1018 XE Amsterdam, The Netherlands}


\date{\today}

\begin{abstract}
We have observed St\"{u}ckelberg oscillations in the dipole-dipole interaction between Rydberg atoms with an externally applied radio-frequency field. The oscillating RF field brings the interaction between cold Rydberg atoms in two separated volumes into resonance. We observe multi-photon transitions when varying the amplitude of the RF-field and the static electric field offset. The angular momentum states we use show a quadratic Stark shift, which leads to a fundamentally different behavior than linearly shifting states. Both cases are studied theoretically using the Floquet approach and are compared. The amplitude of the sidebands, related to the interaction strength, is given by the Bessel function in the linearly shifting case and by the generalized Bessel function in the quadratically shifting case. The oscillatory behavior of both functions corresponds to St\"{u}ckelberg oscillations, an interference effect described by the semi-classical Landau-Zener-St\"{u}ckelberg model. The measurements prove coherent dipole-dipole interaction during at least 0.6~$\mu$s. 
\end{abstract}


\maketitle

\section{Introduction}

The subject of the sinusoidal perturbation of a quantum mechanical object is a very general one and has a long history. Here we will treat this subject for the particular case of particles with states that have a quadratic shift under the influence of a perturbation. This is what is generally found in second-order perturbation theory. The typical example is the polarizability of a neutral particle under the influence of an electric field. The realization chosen to study this phenomenon is the dipole-dipole interaction between Rydberg atoms, because this system provides the resolution and flexibility required for these experiments. Apart from representing a much broader class of phenomena, the study of dipole-dipole interaction in oscillating fields provides us a direct tool to measure the coherence time of the interaction and the possibility of switching the interaction. Both aspects are relevant for the feasibility of implementing quantum information processing with Rydberg atoms. 


 
First we will give a description of polar and polarizable systems in a RF field, i.e. we compare the situation for states with a linear versus a quadratic Stark shift. We use the Floquet approach, which is applicable for periodically varying fields. In this framework a two-level system plus oscillating field is replaced by an infinite number of sidebands on the two states. For a quadratic Stark shift the result is fundamentally different and much richer than for a linearly Stark shifted state. Next we compare the Floquet description with the classical limit. In addition we interpret the oscillations of the population of the sidebands as St\"{u}ckelberg oscillations. 

In the subsequent section we describe details of the experiment and present results of the measurements. Related experiments have been performed with transitions within single Rydberg atoms, in most cases for states with a linear Stark shift \cite{PRL.57.2512,PRL.74.355,PRL.77.2424}, but also quadratic shifts have been studied \cite{PRA.50.1101}. Collisions of Rydberg atoms in an oscillating field have been studied in an atomic beam setup \cite{PRL.67.2287,PRL.49.191}, as well as in a cold cloud \cite{PRL.98.203005,PRA.65.063404}. In our experiment we control the interaction strength by maintaining a fixed distance between the Rydberg-atom volumes. Because of the small unperturbed energy mismatch between initial and final states, we use a rather low frequency for the oscillating field; it lies in the radio frequency range rather than in the microwave range, giving the opportunity to test the coherence properties of the system for a longer period of time.


\section{A coupled two-level system in an oscillating field}\label{sec:theo}

To introduce the system under investigation we recall the two-level system \cite{tombook}, coupled by an interaction $V$ in the presence of a field $F(t)$
\begin{equation}\label{eq:ft}
F(t)=F_S + F_{RF} \cos \omega t ,
\end{equation}
with an oscillating and a static part. We write the full Hamiltonian as
\begin{equation}
H = H_0 + H_{F_S} + H_{F_{RF}} + V ,
\end{equation}
where $H_0$ is the Hamiltonian of the two-level system without interaction, $H_{F_S}$ the interaction with the static field and $H_{F_{RF}}$ the interaction with the oscillating field. V is the dipole interaction between the two involved states. We want to know the population in both states, so we need to know the full wavefunction $\psi(\textbf{r},t)$, by solving the Schr\"{o}dinger equation. We solve the problem by starting from the known static solutions for $H_0+H_{F_S}$, as described below, and then we add the other two terms as perturbation. In the next sub-section \ref{sec:num} `Numerical solutions in the time domain' we work out the problem by first adding the interaction term $V$ to the Hamiltonian, and then the oscillating field $H_{F_{RF}}$. This provides us with two coupled equations which can be solved numerically in the time domain. In the subsequent sub-sections we work in the reverse order. In section \ref{sec:floquet} `Floquet approach' we add the oscillating field $H_{F_{RF}}$ first and work out the problem in the frequency domain, which basically provides us with an infinite number of steady states in the form of sidebands. The classical limit of the sideband population, section \ref{sec:sideband}, gives some physical insight in its behavior. Then, in section \ref{sec:coupling} `Coupling between dressed and undressed state' we add the interaction term $V$ to the Hamiltonian. The oscillations in the coupling as a function of RF amplitude can be understood in terms of St\"{u}ckelberg oscillations, explained in section \ref{sec:stueck}. 

We will start by treating the perturbation of the Hamiltonian $H_0$ in a static field. With just the first two terms of the Hamiltonian $H=H_0+H_{F_S}$ we obtain the solution
\begin{equation}
\psi(\textbf{r},t) = a\, \psi_1(\textbf{r},t) + b\, \psi_2(\textbf{r},t)\label{eq:wavefun} ,
\end{equation}
where $\psi_1$ and $\psi_2$ are the eigenstates of the system and $a$ and $b$ are their normalized time-independent amplitudes,
\begin{equation}
    \psi_{1 (2)}(\textbf{r},t) = \psi_{1 (2)}(\textbf{r}) e^{-i W_{1 (2)} t} ,
\end{equation}
where $W_1$ and $W_2$ are the eigenenergies of the two stationary states.

\begin{figure}[htb]
\begin{minipage}[t]{.4\textwidth}
\includegraphics[width=1\textwidth]{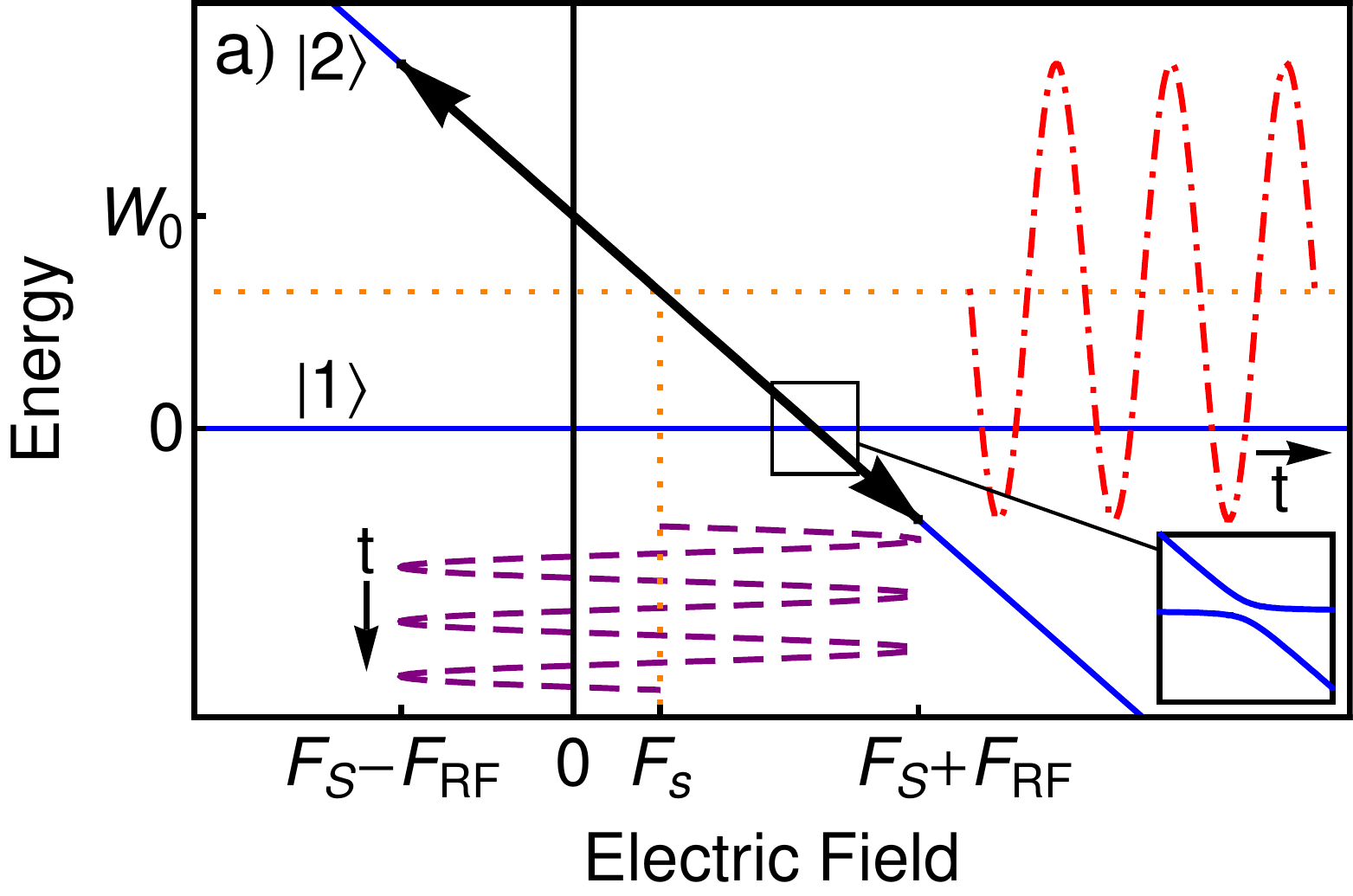}
\end{minipage}
\hfill
\begin{minipage}[t]{.4\textwidth}
\includegraphics[width=1\textwidth]{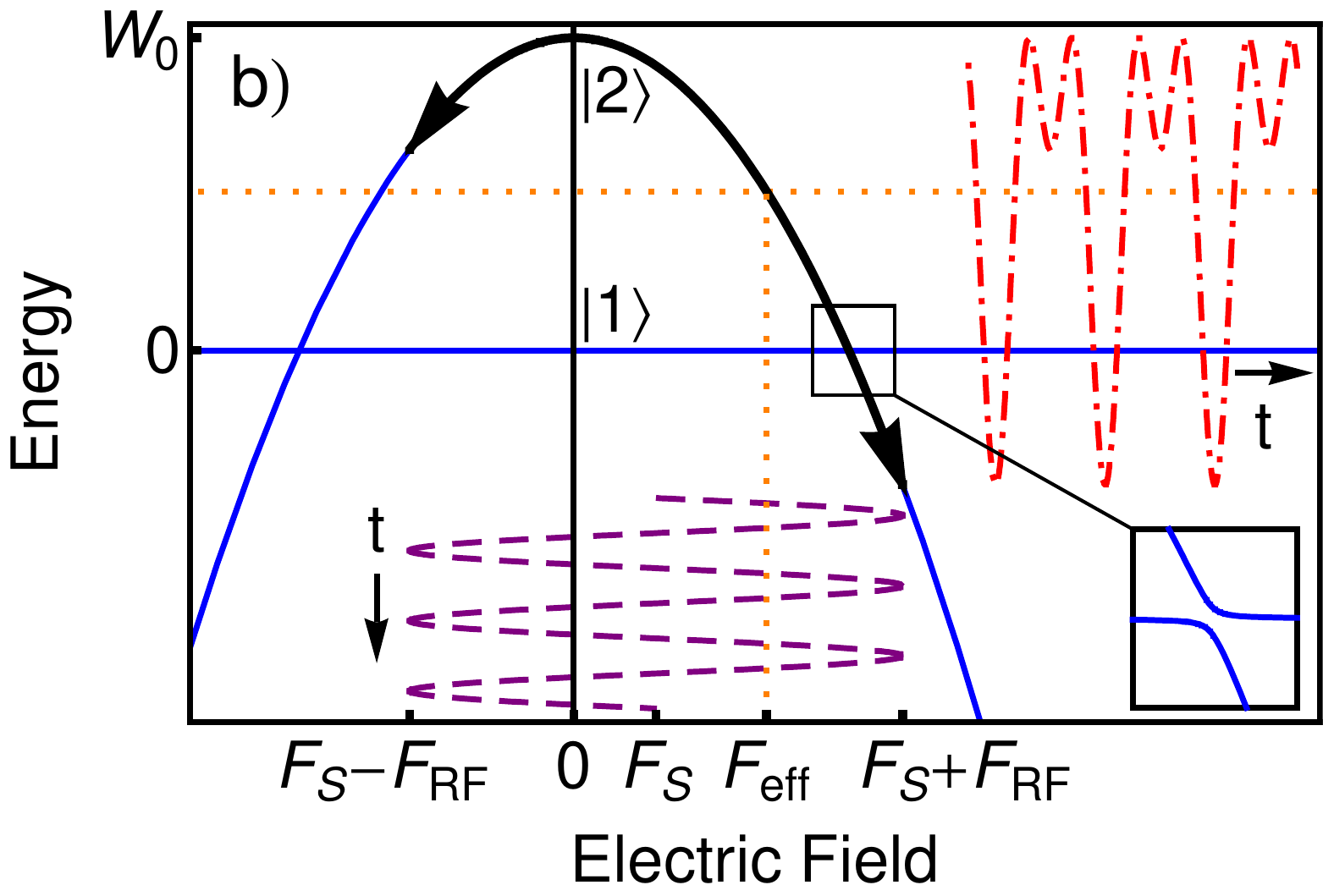}
\end{minipage}
\caption{(Color online): Energy level diagrams of state $|1\rangle$ and $|2\rangle$ as a function of electric field. State $|1\rangle$ has no Stark shift, state $|2\rangle$ has a linear Stark shift in Fig.~a) and a quadratic Stark shift in Fig.~b). Due to the coupling between the states, the crossing becomes an avoided crossing, depicted in the inset. When an oscillating field is applied, depicted as a dashed purple line, this translates as an oscillating energy for state $|2\rangle$, depicted on the right hand side as a dash-dotted red line. The orange dotted lines depict the average energy during the oscillation and the corresponding field; in a) this field is simply $F_S$, the average field, but in b) it is different and we define it as the effective field $F_{\mathrm{eff}}$, given in Eq.~\ref{eq:feff}.}\label{fig:introstuck}
\end{figure}

We will discuss two cases, one for states with a linear Stark shift and one for states with a quadratic Stark shift. The first case is depicted in Fig.~\ref{fig:introstuck}a. State $|1\rangle$ always has the energy 0 and state $|2\rangle$ has the energy $W_0$ in the absence of any fields and a linear energy shift as a function of electric field $F$ (all equations are in atomic units):
\begin{equation}
 W_1 = 0; \qquad W_2 = W_0 - k F .
\end{equation}
Here $k$ is the permanent dipole moment of state $|2\rangle$. This system is in fact equivalent to two states with difference energy $W_0$ and a difference in dipole moment of $k$, which makes this two-level system applicable to many realistic systems. The two levels cross at a field
\begin{equation}\label{eq:fl}
F_{\mathrm{lin}} = \frac{W_0}{k} .
\end{equation}

The second case is sketched in Fig.~\ref{fig:introstuck}b. Again, state $|1\rangle$ has the energy 0 for all fields and state $|2\rangle$ has the energy $W_0$ at zero field but the energy decreases quadratically as a function of electric field $F$:
\begin{equation} 
W_1 = 0; \qquad W_2 = W_0 - \frac{1}{2} \alpha F^2 .
\end{equation} 
Here $\alpha$ is the polarizability of state $|2\rangle$. Similar to the linear case this system is equivalent to two states with difference energy $W_0$ and a difference in polarizability of $\alpha$. The two levels cross at $\pm F_{\mathrm{quad}}$, where
\begin{equation}\label{eq:fq}
F_{\mathrm{quad}} = \sqrt{\frac{2 W_0}{\alpha}} .
\end{equation}


\subsection{Numerical solutions in the time domain}\label{sec:num}
The most direct method to calculate the evolution of the wavefunction in an oscillating field is to numerically solve the two coupled differential equations for both states. For this we first add the coupling $V$ and then the time dependent field perturbation $H_{F_{RF}}$. When we add the interaction $V$, the degeneracy at the crossing is lifted and an avoided crossing arises, which has a width $\Omega_0$. This $\Omega_0$ can also be interpreted as a quantum beat oscillation frequency: the population of state $\psi_1$ and $\psi_2$ oscillate against each other with this frequency after an appropriate initial condition. The avoided crossing is depicted in the inset of Fig.~\ref{fig:introstuck} where the width is given by
\begin{equation}\label{eq:avcross}
\Omega_0 = 2 \langle
\psi_2(\mathbf{r})|V|\psi_1(\mathbf{r})\rangle .
\end{equation}
When we add a time-dependent field it is convenient to separate the time and spatial dependence of the wavefunction (Eq.~\ref{eq:wavefun})
\begin{equation}
\psi(\textbf{r},t) = T_1(t) \, \psi_1(\textbf{r}) + T_2(t) \,  \psi_2(\textbf{r})\label{eq:wavefun2} .
\end{equation}
where $\psi_1(\textbf{r})$ and $\psi_2(\textbf{r})$ are the two original states without interaction. This separation of time- and spatial dependence is only allowed if both spatial states remain approximately the same despite the applied fields. We will see in section \ref{sec:spectr} that this assumption is valid. Using this wavefunction in the time-dependent Schr\"{o}dinger equation we obtain
\begin{equation}\label{eq:twocoupled} 
i \, \dot{T}_{1 (2)}(t) = W_{1 (2)}(t) + \frac{\Omega_0}{2} T_{2 (1)}(t).
\end{equation}
Now the energies $W_1$ and $W_2$ depend on time. For the linear case we have
\begin{equation}\label{eq:en_time_lin}
    W_2(t) = W_0 - k \left(F_S + F_{RF} \cos \omega t\right) ,
\end{equation}
and for the quadratic case 
\begin{equation}\label{eq:en_time_quad}
W_2(t) = W_0 - \frac{1}{2} \alpha \left(F_S^2 + 2 F_S F_{RF} \cos
\omega t  + F_{RF}^2 \cos^2 \omega t \right)
\end{equation}
and in both cases $W_1(t)\!=\!0$.

We have numerically solved the two coupled equations in Eq.~\ref{eq:twocoupled}. However, the results are less informative than the approach described in the next section, the Floquet approach. Although equivalent, an answer in terms of a coherent superposition of stationary states is more informative than the time-dependence of a wavefunction. However, this numerical approach can be used for all time-dependent fields, whereas the Floquet approach only works for periodic functions.


\subsection{Floquet approach}\label{sec:floquet}

In the Floquet approach we add the perturbations in the opposite order as in the numerical calculation; so we first perturb the states $\psi_1$ and $\psi_2$ with the oscillating field and subsequently add the interaction $V$ between the two states \cite{tombook,PRA.50.1101}. As in the previous section we separate the time and spatial dependence, but this time we keep two separate wavefunctions
\begin{equation}\label{eq:2_separate_wavefun} 
\psi_{1 (2)}(\textbf{r},t) = T_{1 (2)}(t) \, \psi_{1 (2)}(\textbf{r}).
\end{equation} 
Using these in the time-dependent Schr\"{o}dinger equation we obtain
\begin{equation}\label{eq:tdse_12} 
    i \, \dot{T}_{1 (2)}(t) \, \psi_{1 (2)}(\textbf{r}) = W_{1 (2)}(t) \, T_{1 (2)}(t) \, \psi_{1 (2)}(\textbf{r}) .
\end{equation} 
We can factor out the spatial dependent part and integrate Eq.~\ref{eq:tdse_12}
\begin{equation} 
    T_{1(2)}(t) = e^{-i \int W_{1(2)}(t)\,dt}\label{eq:t2int} ,
\end{equation} 
where $W_1(t) \!=\! 0$ for both examples.

\begin{figure}[htb]
\includegraphics[width=.48\textwidth]{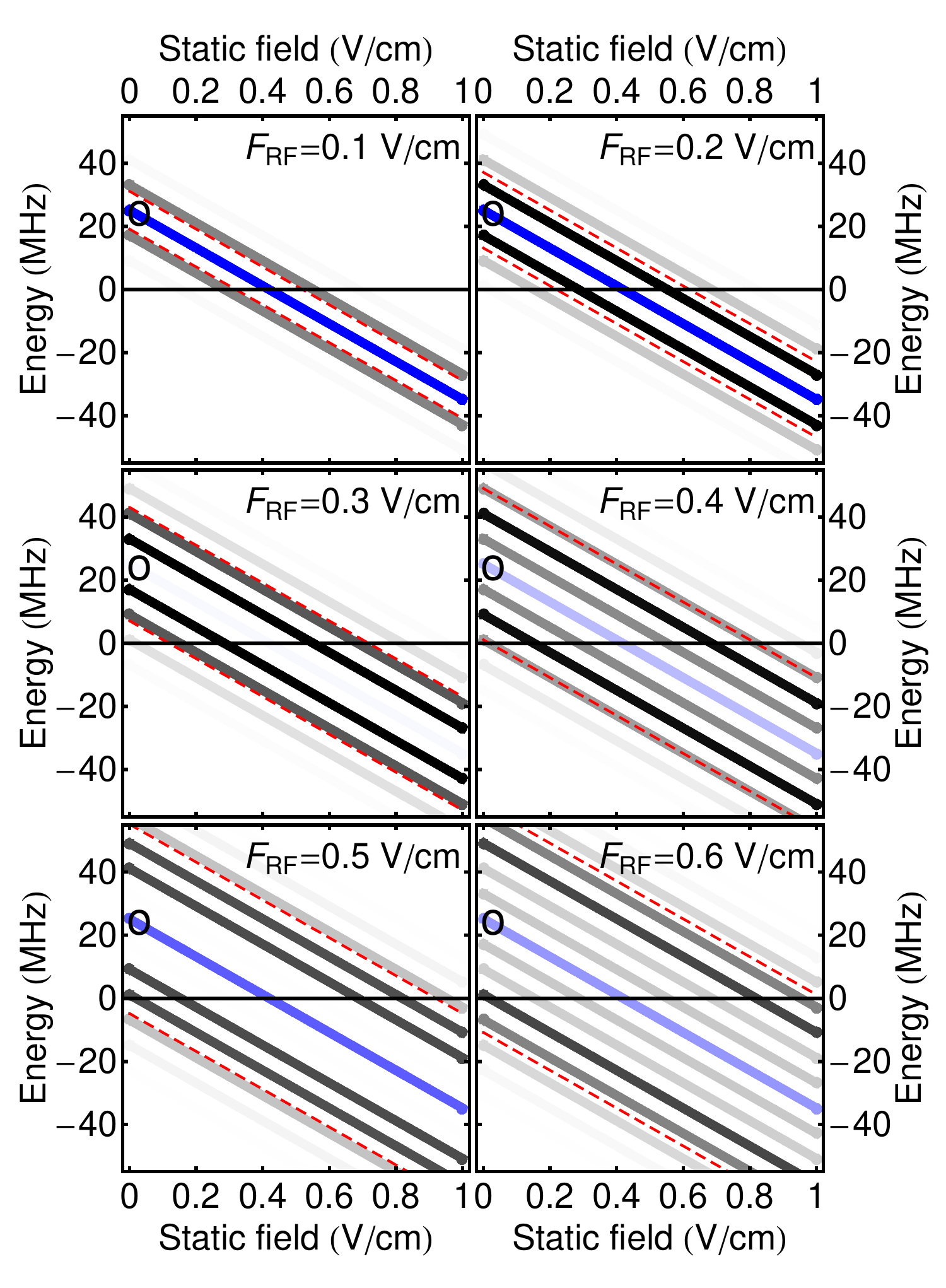}
\caption{(Color online)The sidebands of state $|2\rangle$ for the linear case in an oscillating field of 8~MHz for various RF amplitudes. The intensity of a line represents the population of the respective sideband, where the original state, or the $n=0$ sideband, is colored blue and indicated with an ''o''. The red dashed lines depict the borders of the classically allowed region. We used $k=2 \pi \times 60$~MHz/(V/cm) and $W_0=2 \pi \times 25$~MHz.} \label{fig:sidebandslin}
\end{figure}

For the linearly Stark shifted case, still following ~\cite{tombook}, we use Eq.~\ref{eq:en_time_lin} to obtain
\begin{equation}
T_2(t) = e^{-i\left(W_0 - k F_S \right) t} e^{i \frac{k
F_{RF}}{\omega} \sin\omega t} .
\end{equation}
The last exponent can be written as an expansion of Bessel functions \cite{abramowitz}, such that
\begin{equation}\label{eq:wavefunlin}
\psi_2(\textbf{r},t) = \psi_2(\textbf{r}) \, e^{-i\left(W_0 - k
F_S \right)t} \sum_{n=-\infty}^{\infty} J_n\left(\frac{k
F_{RF}}{\omega}\right) e^{i n \omega t} .
\end{equation}
This expression shows that the wavefunction consists of the original spatial wavefunction, but with a modified time dependence in the form of an infinite number of sidebands. This is the steady-state solution of the problem, consisting of a number of states with different amplitudes, given by the Bessel function $J_n\left(\frac{k F_{RF}}{\omega}\right)$, all with a different energy
\begin{equation}
  W_{2,n} = W_0 - k F_S - n \omega  \label{eq:en_sidebands_lin}.
\end{equation}
\begin{figure}[htb]
\includegraphics[width=.48\textwidth]{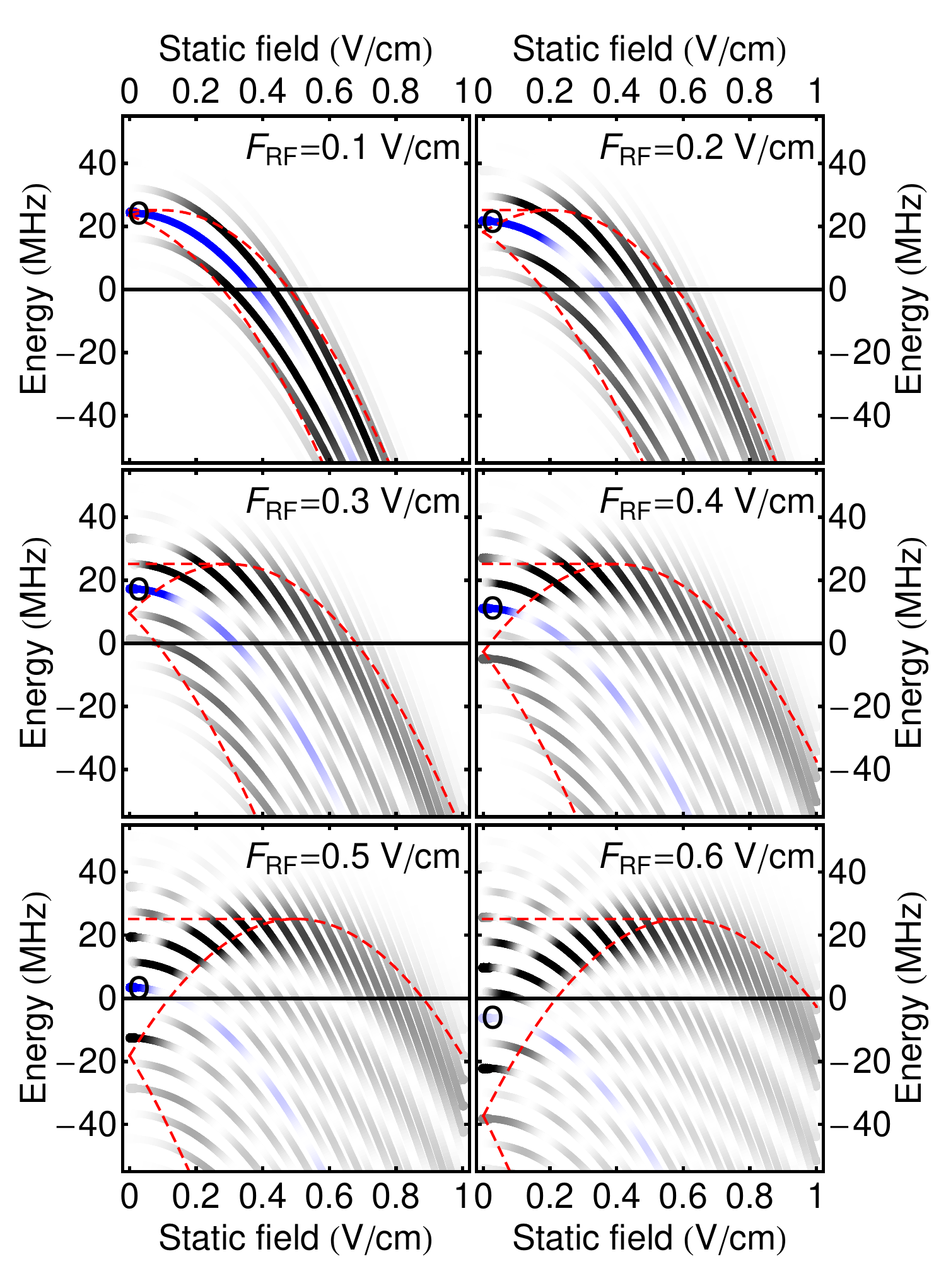}
\caption{(Color online) The sidebands of state $|2\rangle$ for the quadratic case case in an oscillating field of 8~MHz for different RF amplitudes. The intensity of a line symbolizes the population of that sideband. The $n=0$ sideband is depicted in blue and indicated with an ''o''. The red dashed lines depict the position of the asymptotes of the distribution of occurring energies. We used the values corresponding to the experimental values of the left resonance, as mentioned at Eq.~\ref{eq:reaction1}.} \label{fig:sidebandsquad}
\end{figure}
In Fig.~\ref{fig:sidebandslin} the sidebands of $\psi_2$ are plotted for various RF amplitudes. It is clearly visible that more sidebands are populated when a larger RF amplitude is applied. The red dashed lines indicate the classical limits, the maximum and minimum field values that occur in the oscillation. This is equivalent with the absolute order and absolute argument of the Bessel function being equal.

The result for the quadratically Stark shifted case is quite different. Combining Eqs.~\ref{eq:t2int} and \ref{eq:en_time_quad} we obtain
\begin{align}
T_2(t) &= e^{-i\left(W_0 - \frac{1}{2} \alpha \left(F_S^2 +\frac{1}{2}F_{RF}^2\right)\right) t} \nonumber \\ 
&\times e^{i\frac{\alpha F_{RF} F_S}{\omega} \sin \omega t} e^{i\frac{\alpha F_{RF}^2}{8 \omega} \sin 2 \omega t} .
\end{align}
Here we have two exponents to expand in terms of Bessel functions
\begin{align}
 e^{i x \sin \omega t} &e^{i y \sin 2 \omega t} = {}  \nonumber \\
&= \sum_{m'=-\infty}^{\infty} J_{m'}(x) e^{i m' \omega t}\sum_{m=-\infty}^{\infty} J_m(y) e^{i 2 m \omega t} \nonumber \\
&= \sum_{n=-\infty}^{\infty} e^{i n \omega t} \left(\sum_{m=-\infty}^{\infty} J_{n-2m}(x) J_m(y)\right),
\end{align}
where $n=m'+2m$ is used on the third line. The amplitude of the sideband, i.e. the inner summation in the above expression, is known as the generalized Bessel function $J_n(x,y)$. It was first used by Reiss \cite{Reiss1962} and investigated by e.g.\ Dattoli et al.~\cite{G.Dattoli1990}. It is defined here as
\begin{equation}
J_n(x,y) = \sum_{m=-\infty}^{\infty} J_{n-2m}(x) J_m(y)\label{eq:gn} .
\end{equation}
In practice the summation can be constrained to $|m|\leq y+3 y^{1/3}+3$ or $|m|\leq\frac{1}{2}(n+x+3 x^{1/3}+3)$, because beyond these points the Bessel function is negligibly small \cite{explain}. Instead of evaluating the sum, it is also possible to calculate the generalized Bessel function on basis of the recurrence relation \cite{lotstedt2009} in analogy with the well-known approach to calculate the regular Bessel function on basis of its recurrence relation. 

Using these results the wavefunction becomes
\begin{align}\label{eq:wavefunquad}
\psi_2(\textbf{r},t) &= \psi_2(\textbf{r}) e^{-i\left(W_0 -\frac{1}{2}\alpha \left(F_S^2 + \frac{1}{2} F_{RF}^2 \right) \right)t} \nonumber \\
 &\times \sum_{n=-\infty}^{\infty} e^{i n \omega t} J_n\left(\frac{\alpha F_{RF} F_S}{\omega} ,  \frac{\alpha F_{RF}^2}{8 \omega}\right).
\end{align}
A similar description of sidebands on quadratically shifted states is given in \cite{PRA.50.1101}. Note that the leading order of the series expansion is proportional to $F_{RF}^n$; a result that is also found for the linear case and that is expected from a description of multi-photon transitions.

If we compare this wavefunction with the linear case it is the generalized Bessel function instead of the regular Bessel function that determines the sideband amplitude. The result depends on both the static field $F_S$ as well as the RF amplitude $F_{RF}$, whereas for the linear case it depends on the RF amplitude only. The energy of the sidebands is also fundamentally different and one obtains
\begin{equation}
  W_{2,n} = W_0 - \frac{1}{2}\alpha \left(F_S^2 + \frac{1}{2} F_{RF}^2
  \right) - n \omega \label{eq:en_sidebands_quad} .
\end{equation}
Whereas in the linear case the energy of a particular sideband shifts with the static field $F_S$ only, in the quadratic case there is an additional term that depends on the RF amplitude $\frac{1}{4} \alpha F_{RF}^2$. This term corresponds to the AC-Stark shift of the state. It is convenient to define an effective field $F_{\mathrm{eff}}$, such that the energy shift of such a static field corresponds to the energy shift of a given static plus RF field:
\begin{equation}\label{eq:feff}
  F_{\mathrm{eff}}^2 = F_S^2 + \frac{1}{2} F_{RF}^2 .
\end{equation}
The different behavior is visible in Fig.~\ref{fig:sidebandsquad}, where the sidebands are plotted versus static field for several RF amplitudes. Again, the intensity depicts the population of the sideband: $J_n\left(\alpha F_{RF} F_S/\omega,\alpha F_{RF}^2/(8 \omega)\right)^2$. As in the linear case more sidebands are populated with a larger RF field, but a clear difference is that the population of a sideband is now not only a function of the RF amplitude, but also of the DC field, visible as the changing intensity within one sideband. Secondly, the different positions of the sidebands in each subfigure illustrates that the sideband energy also depends on $F_{RF}$ (note that the original state, or the $n\!=\!0$ sideband, in blue and indicated with ''o'', is shifted downwards for larger $F_{RF}$). 

The next step would be to add the coupling between state $|1\rangle$ and state $|2\rangle$ (section \ref{sec:coupling}), but first we will look closer into the population of the sidebands for different frequencies.


\subsection{Classical limit of the sideband population distribution}\label{sec:sideband}

\begin{figure}[htb]
\includegraphics[width=.48\textwidth]{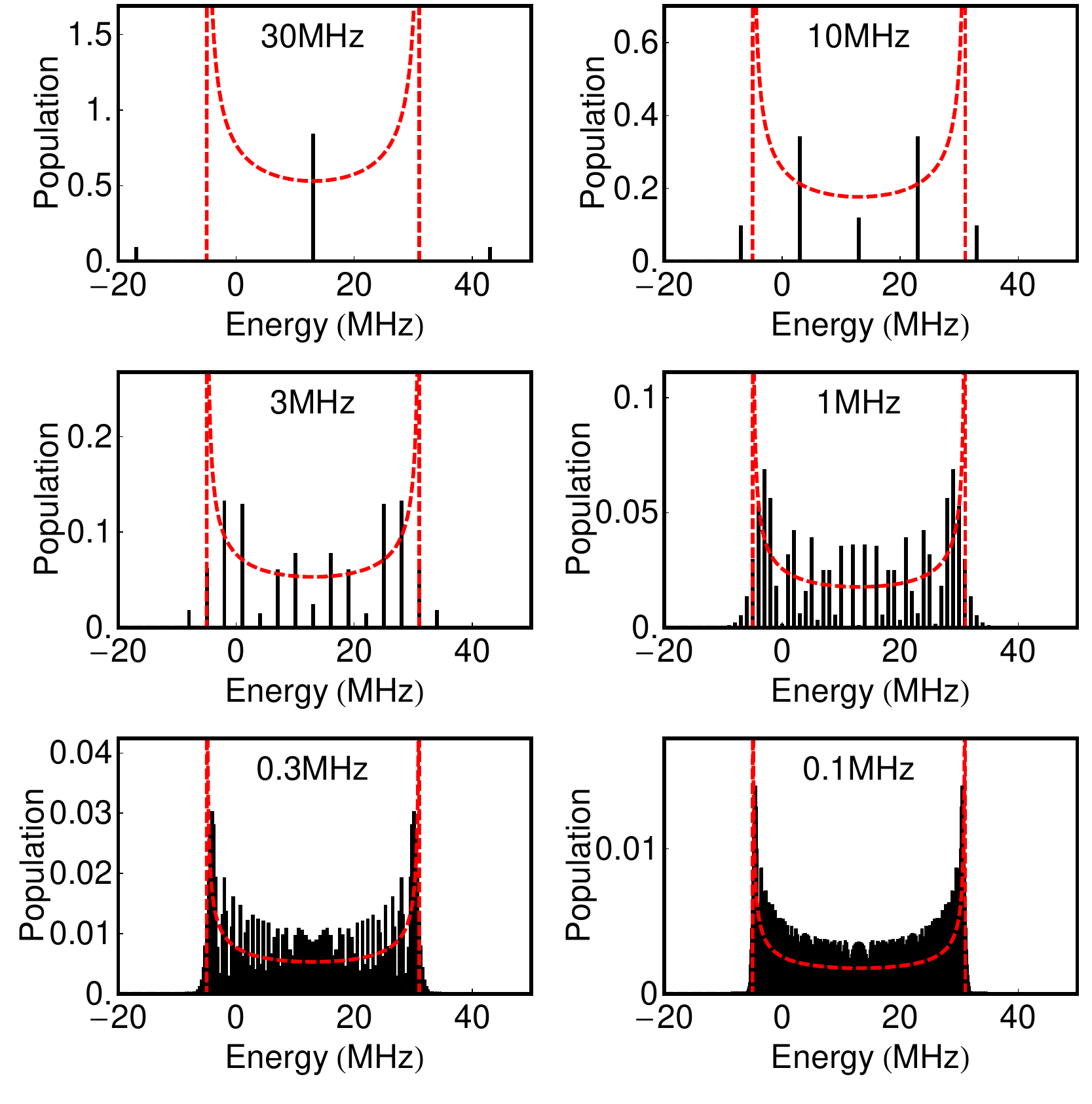}
\caption{(Color online) Depicted in black is the sideband population of state $|2\rangle$ as a function of the sideband energy (Eqs.\ \ref{eq:en_sidebands_lin} and \ref{eq:en_sidebands_quad}) for various frequencies. $\omega$ is $2\pi$ times the denoted frequency. We used again $k=2 \pi \times 60$~MHz/(V/cm) and $W_0=2 \pi \times 25$~MHz and field values of $F_S=0.2$~V/cm and $F_{RF}= 0.3$~V/cm. The red dashed lines depict the distribution of clasically occurring energies, calculated with the approach of Eq.~\ref{eq:densfun2}.} \label{fig:classbessel}
\end{figure}

\begin{figure}[htb]
\includegraphics[width=.48\textwidth]{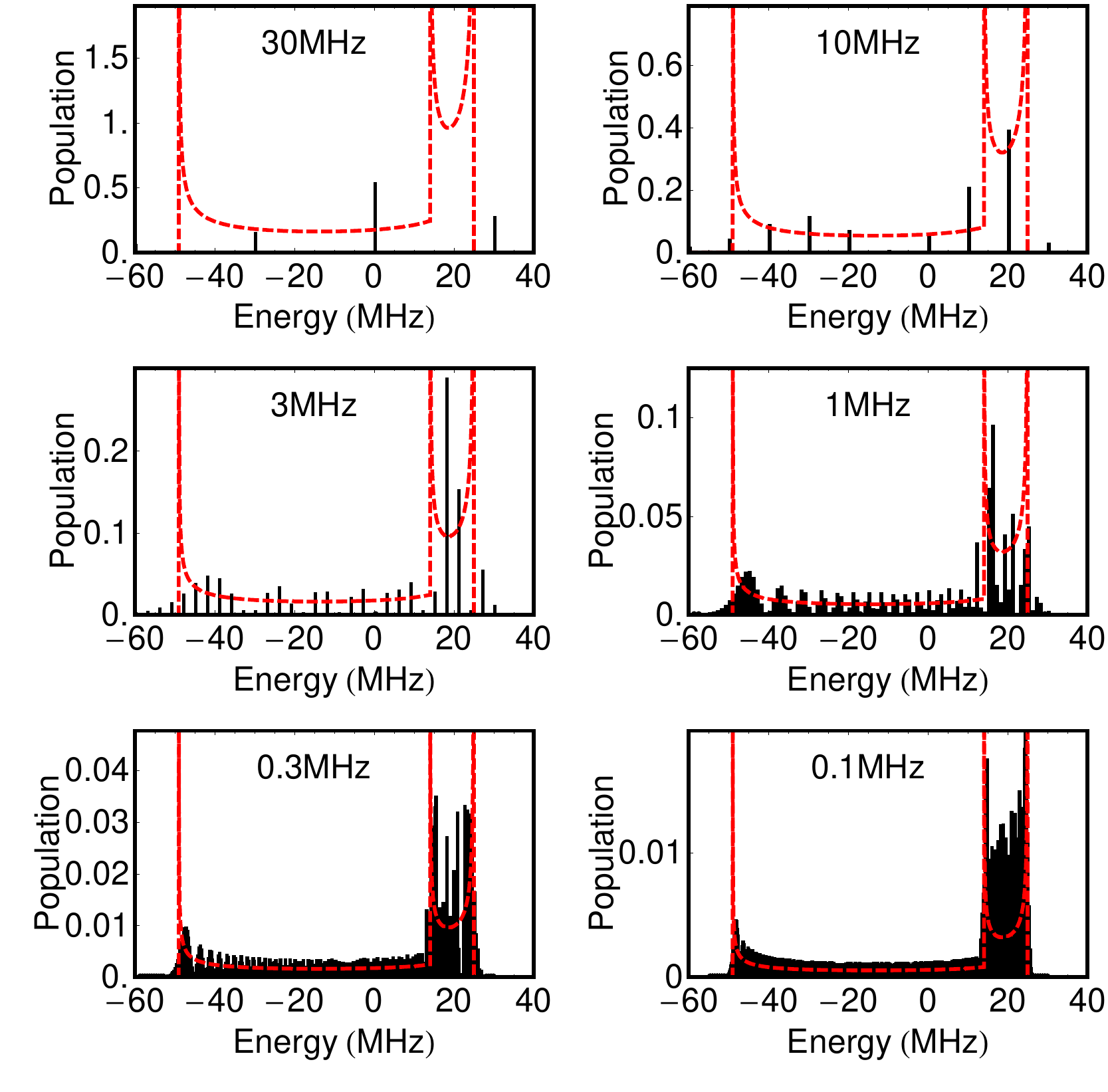}
\caption{(Color online) Similar as Fig.\ \ref{fig:classbessel}. We used the experimental values for the left resonance (Eq.\ \ref{eq:reaction1}) and field values of $F_S=$0.2~V/cm and $F_{RF}=$0.45~V/cm. The distribution of occurring energies (red dashed lines) is calculated with Eq.~\ref{eq:densfun2}. Note the occurence of three turning points.} \label{fig:classzhang}
\end{figure}

To illustrate the sideband population and to obtain some physical insight into its behavior, we depicted the sideband population for a fixed field amplitude and offset as a function of the sideband energy (Eqs.\ \ref{eq:en_sidebands_lin} and \ref{eq:en_sidebands_quad}) and various frequencies in Figs.~\ref{fig:classbessel} and \ref{fig:classzhang}. So we plot the squared (generalized) Bessel function $J_n(x)^2$ respectively $J_n(x,y)^2$ for fixed $x$ and $y$ (per subfigure) as a function of $n$. The pictures can also be seen as a vertical cut through a subfigure of Fig.~\ref{fig:sidebandslin} or \ref{fig:sidebandsquad} with the appropriate parameters $F_S$, $F_{RF}$ and $\omega$. For large frequencies, i.e.\ small $x$ and $y$, the function peaks at $n=0$ and drops off on both sides, so there are not many populated sidebands. For smaller frequencies (or large $x$,$y$), more and more sidebands arise. Around $n\!=\!0$ the population oscillates and near $|n|=|x|$ the Bessel function reaches a maximum, after which it drops exponentially to zero. Interestingly, the sideband populations extend further in energy beyond $|n|>|x|$ for the higher frequencies as for the lower frequencies. For the number of populated sidebands beyond $|n|=|x|$ the opposite is true. These extensions can be understood as a quantum mechanical effect, whereas the energy levels are in a classically forbidden region. The generalized Bessel function also has these features. In addition, it has a more asymmetric structure, which we will discuss further below.

For low frequencies the (generalized) Bessel function approaches the classical limit, meaning it approaches the density distribution of the occurring energies. To calculate such a density distribution it is useful to know how a distribution, $f(x)$, changes under a change of variable, from $x$ to $y$. If $y=g(x)$ the resulting density function becomes
\begin{equation}\label{eq:densfun2}
P(y) = \sum_k^n\left| \frac{1}{g'(g_k^{-1}(y))}\right|
f(g_k^{-1}(y))
\end{equation}
with $g'$ the derivative and $g^{-1}$ the inverse function and $n$ the number of solutions in $x$ for $g(x)=y$. $g_k^{-1}(y)$ are these solutions.

\begin{figure}[htb]
 \includegraphics[width=.32\textwidth]{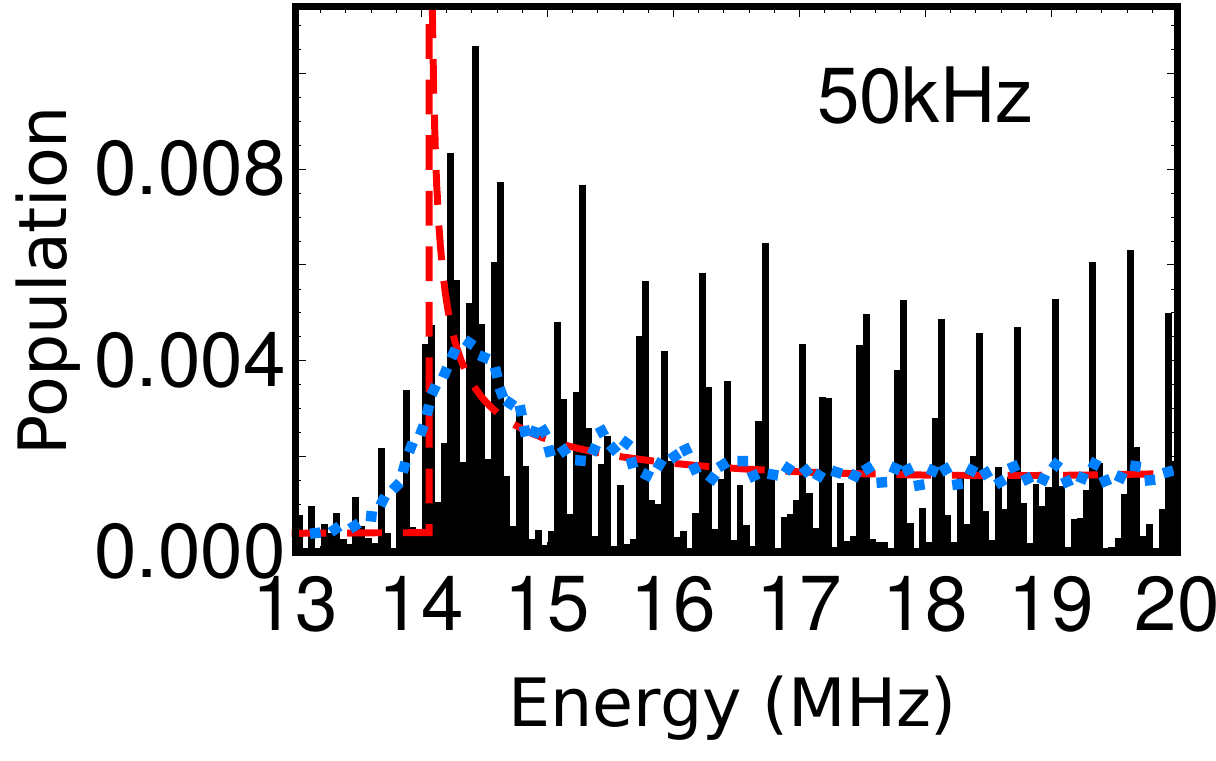}
\caption{(Color online) As Fig.~\ref{fig:classzhang}, but at a lower frequency and zoomed in around the middle asymptote. The blue dotted curve depicts the moving average of the sideband population depicted in black.} \label{fig:movingavg}
\end{figure}

We could choose $g$ to be the energy as a function of time, given by Eq.~\ref{eq:en_time_lin} and \ref{eq:en_time_quad}. It is, however, instructive and easier to first calculate the distribution of the occurring fields. So we use $g(t) = F(t) = F_S+F_{RF}\cos \omega t$. We start with $f$ as a uniform normalized distribution. It is sufficient to have instances during half a cycle of the cosine, 
\begin{equation}
f(t) = \left\{ \begin{array}{ll}
\frac{\omega}{\pi} & \textrm{if $0 \le \omega t \le \pi$}\\
0 & \textrm{elsewhere}
\end{array} \right.
\end{equation}
Application of the functions $f$ and $g$ to Eq.~\ref{eq:densfun2} provides the probability density distribution of the field
\begin{align}\label{eq:densfield}
&P(F) = \nonumber \\ 
&\left\{\! \begin{array}{ll}
\frac{1}{\pi \sqrt{F_S^2-(F-F_{RF})^2}} & \textrm{if $F_S\!-\!F_{RF}\! \le \!F \!\le \!F_S\!+\!F_{RF}$}\\
0 & \textrm{elsewhere}
\end{array} \right.
\end{align}
This function diverges at $F=F_S \pm F_{RF}$. The asymptotes correspond to the turning points of the field.

The function $F(t)$ is sketched in the energy level diagram Fig.~\ref{fig:introstuck}, as a dashed purple line. How the field oscillation translates to an oscillation in energy (Eq.~\ref{eq:en_time_lin} and \ref{eq:en_time_quad}) is depicted on the right hand side as the red dash-dotted line. In the linear case, the function simply remains a cosine, but in the quadratic case the function is distorted. For $F_{RF}>F_S$ the density distribution consists of two parts that add up.  In this case there is a third turning point when the field goes through zero. Both functions are plotted together with the strength of the sidebands in Figs.~\ref{fig:classbessel} and \ref{fig:classzhang}. Here they are multiplied by the frequency to match the average height of the sidebands. The asymptotes of the distribution functions are also depicted in Figs.~\ref{fig:sidebandslin} and \ref{fig:sidebandsquad} as red dashed lines.

We observe in Figs.~\ref{fig:classbessel} and \ref{fig:classzhang} that for the lower frequencies the sideband population indeed converges nicely to the distribution functions. The height of the peaks oscillates around the red dashed line. In most cases values occur between 0 and twice the red dashed line. In the case of the generalized Bessel function, on the right hand part of the plot some much higher peaks occur. However, the moving average still approaches the red dashed line. This is depicted in Fig.~\ref{fig:movingavg}, where we zoomed in around the asymptote $W_0-\frac{1}{2} \alpha (F_S - F_{RF})^2$. 


\subsection{Coupling between dressed and undressed state}\label{sec:coupling}

\begin{figure}[htb]
\includegraphics[width=.48\textwidth]{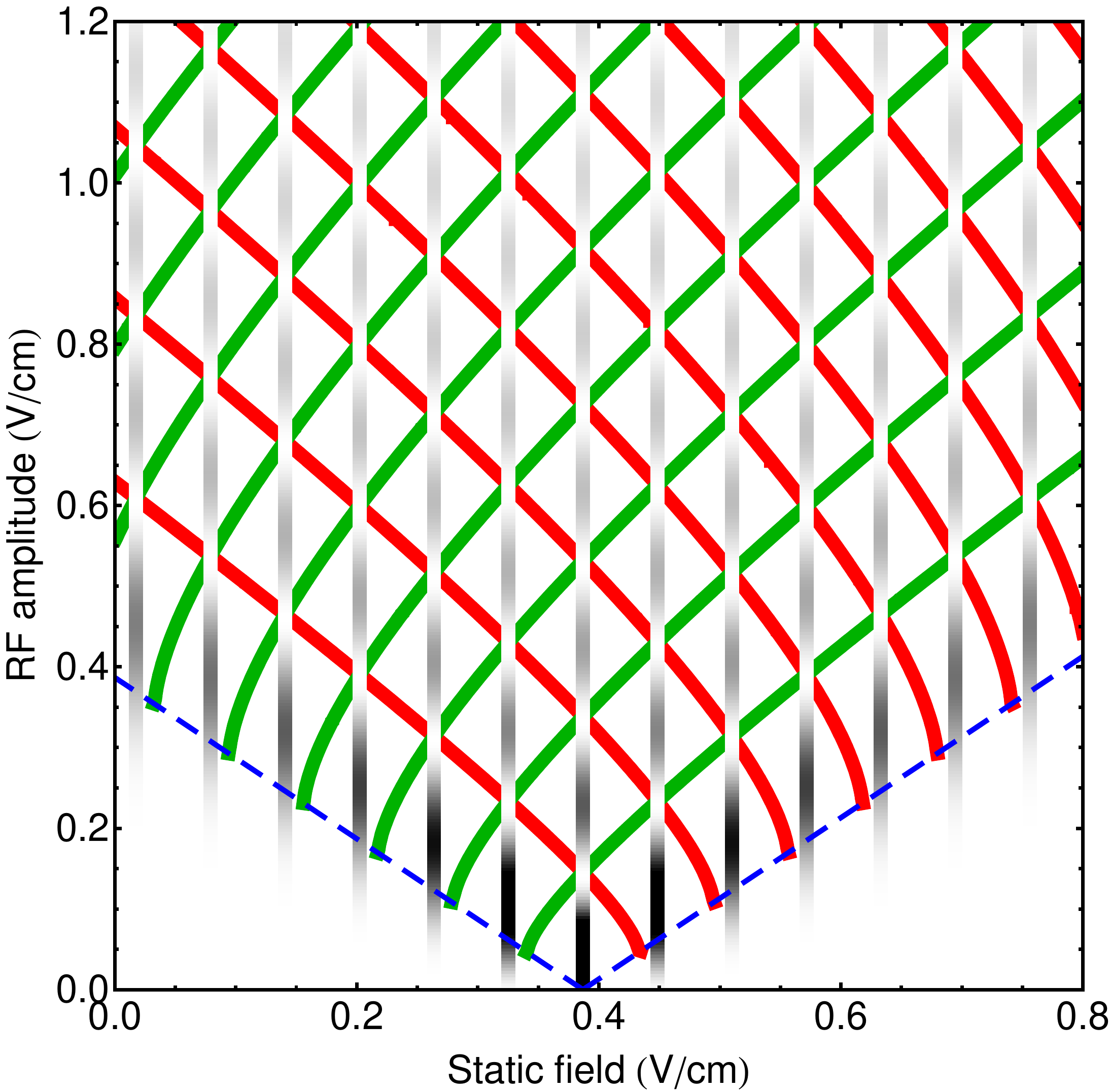}
\caption{(Color online) Calculated interaction strength as a function of static and oscillating field for the linear case. The vertical lines depict the $n\omega$ resonances, where $n$ varies from $-6$ (the one most right) to 6 (the one most left) and $\omega$ is fixed to $2\pi\times$4~MHz. The darker the color the stronger the interaction. Destructive interference between the two states is depicted by the curved lines in red and green, according to St\"{u}ckelberg theory described in subsection \ref{sec:stueck}. The dashed blue lines give the boundaries of the classically allowed region; in other words these lines depict field combinations where the crossing (Eq.~\ref{eq:fl}) occurs at an extremum of the RF oscillation. }\label{fig:floqstucklin}
\end{figure}

\begin{figure}[htb]
\includegraphics[width=.48\textwidth]{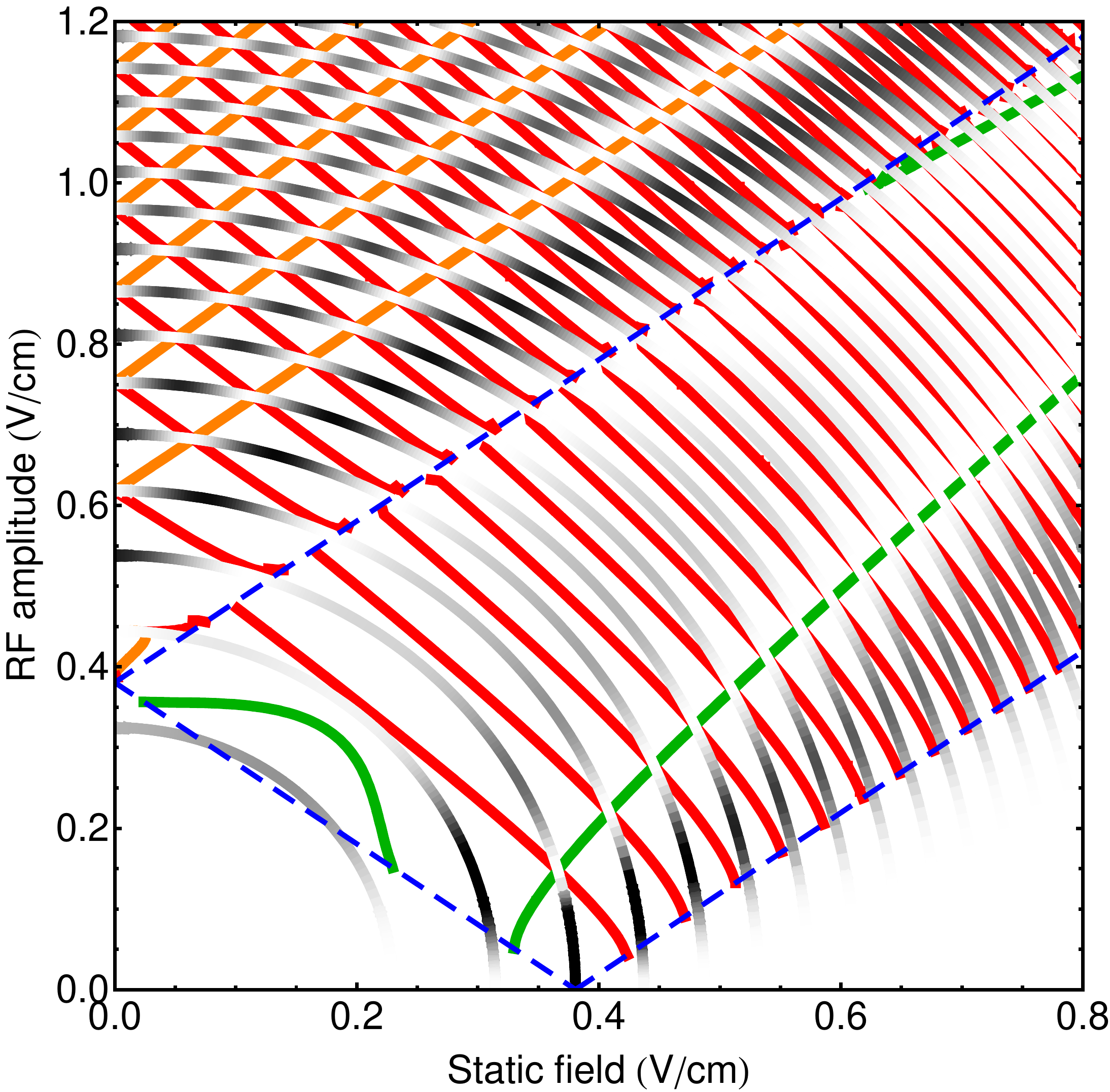}
\caption{(Color online) Calculated interaction strength as a function of static and oscillating field for the quadratic case. The quarter circles in varying gray depict the $n\omega$ resonances. Here $n$ runs from $-25$ (in the upper right corner) to 3 (in the lower left corner); the frequency of the RF field is here $\omega=2\pi\times$8~MHz. The darker the color the stronger the interaction. Destructive St\"{u}ckelberg interference between the two states is depicted by the curved lines in red, green and orange (subsection \ref{sec:stueck}). The dashed blue lines give all field combinations where the crossing is reached during an extremum of the RF oscillation. }\label{fig:floqstuckquad}
\end{figure}

Now that we know the wavefunctions in an oscillating field, we can add the coupling $V$ between $\psi_1$ and $\psi_2$. Every sideband of $\psi_2$ crosses with  $\psi_1$. After introducing the coupling $V$, every crossing becomes an avoided crossing. When we use the wavefunctions given by Eq.~\ref{eq:2_separate_wavefun} we get
\begin{equation}\label{eq:omega}
\left\langle \psi_1(\textbf{r},t) \right| V\left|
\psi_2(\textbf{r},t) \right\rangle = \left\langle
\psi_1(\textbf{r}) \right| V\left| \psi_2(\textbf{r})
\right\rangle T_1^*T_2 .
\end{equation}
The n$^{\textbf{th}}$ sideband of the linear case (see Eq.~\ref{eq:wavefunlin}) becomes
\begin{align}
\left\langle \psi_1(\textbf{r},t) \right| V &\left| \psi_2(\textbf{r},t) \right\rangle_n = \left\langle
\psi_1(\textbf{r}) \right| V \left| \psi_2(\textbf{r})
\right\rangle  \nonumber\\
 & \times e^{-i(W_0 - k F_S - n \omega)t} J_n\left(\frac{k F_{RF}}{\omega}\right)  \label{eq:couplinglin} .
\end{align}
The resonance condition is seen in the exponent
\begin{equation}\label{eq:reslin}
n \omega = W_0 - k F_S .
\end{equation}
The coupling strength at this resonance (the half width of the n$^{\textbf{th}}$ avoided crossing) is given by the time independent part of Eq.~\ref{eq:couplinglin}
\begin{equation}\label{unused}
\frac{\Omega_n}{2} = \left\langle
\psi_1(\textbf{r}) \right| V\left| \psi_2(\textbf{r})
\right\rangle J_n\left(\frac{k F_{RF}}{\omega}\right) 
= \frac{\Omega_0}{2} J_n\left(\frac{k F_{RF}}{\omega}\right) .
\end{equation}
Here we used Eq.~\ref{eq:avcross}, which gives the coupling strength $\Omega_0$ without oscillating field. So the total coupling $\frac{1}{2}\Omega_0$ is distributed over all avoided crossings, according to the Bessel function. In a similar manner, we get the resonance condition for the quadratic case
\begin{equation}\label{eq:resquad}
n \omega = W_0 - \frac{1}{2}\alpha \left(F_S^2 +\frac{1}{2}F_{RF}^2\right)
\end{equation}
and the coupling strength is
\begin{equation}\label{eq:omega_n}
\frac{\Omega_n}{2} = \frac{\Omega_0}{2} J_n\left(\frac{\alpha F_{RF} F_S}{\omega},\frac{\alpha F_{RF}^2}{8 \omega}\right) .
\end{equation}

In Fig.~\ref{fig:floqstucklin} we plot the coupling resonances as a function of RF amplitude, $F_{RF}$, and static field, $F_S$, for a fixed frequency $\omega$ for the linear case. The grayscale depicts the squared coupling strength or $J_n\left(k F_{RF}/\omega\right)^2$. The resonances are just vertical lines, since the resonance depends on the static field only. The blue dashed lines indicate the boundaries of the classically allowed region, which correspond to the regions where the field $F(t)$ (Eq.~\ref{eq:ft}) reaches the static field resonance $F_{\mathrm{lin}}$ (Eq.~\ref{eq:fl}). These classical boundaries correspond to the boundaries depicted as dashed red lines in Fig.~\ref{fig:sidebandslin} and to the asymptotes depicted as red dashed lines in Fig.~\ref{fig:classbessel}. Above these lines the coupling strength oscillates as a function of $F_{RF}$, which we can interpret as St\"{u}ckelberg oscillations, described in the next section.

Fig.~\ref{fig:floqstuckquad} shows the corresponding plot for the quadratic case. Here the resonance lines are curved, in fact, each line follows a line of constant effective field as defined in Eq.~\ref{eq:feff}. The boundaries of the classically allowed region are again indicated with the two lower blue dashed lines. They correspond to the lines $F_S\!+\!F_{RF}\!=\!F_{\mathrm{quad}}$ and $F_S\!-\!F_{RF}\!=\!F_{\mathrm{quad}}$ (with $F_{\mathrm{quad}}$ defined in Eq.~\ref{eq:fq}). There is a third blue dashed line at higher RF amplitudes, which indicates the boundary $F_S\!-\!F_{RF}\!=\!-F_{\mathrm{quad}}$; above this line both the crossing at $-F_{\mathrm{quad}}$ as well as the crossing at $+F_{\mathrm{quad}}$ are involved. Note that the third boundary or asymptote from Figs.~\ref{fig:sidebandsquad} and \ref{fig:classzhang}, the one with value $W_0$, is not relevant, because it is never involved in the crossing with state $|1\rangle$. Within the classically allowed region, i.e.\ above the lower two blue dashed lines, we see the St\"{u}ckelberg oscillations.

At this point there is enough information to start the measurements; it is known where to expect resonances and how strong they are. However, we will first explain why the oscillations in the interaction strength can be interpreted as St\"{u}ckelberg oscillations.


\subsection{St\"{u}ckelberg oscillations}\label{sec:stueck}

St\"{u}ckelberg oscillations \cite{stueckel} are due to an interference effect between two quantum states, which occur when an avoided crossing is traversed at least twice. Imagine we start with all population in the state $|1\rangle$ and subsequently we sweep the field through the resonance and back (Fig.~\ref{fig:introstuck}). When sweeping through, part of the wavefunction will follow the state $|1\rangle$ diabatically and `ignores' the avoided crossing, while the other part will follow the eigen-energy adiabatically and end up in state $|2\rangle$. When coming back again to the crossing, both parts of the wavefunction have built up a different phase, due to their energy difference and due to a difference in passing the crossing. This phase difference leads to interference, which can be observed when varying for example the duration or the amplitude of the field sweep. In this particular case we have a sinusoidal oscillation of the field with a fixed frequency and we vary both $F_S$ and $F_{RF}$. 

\begin{figure}[htb]
\includegraphics[width=0.42\textwidth]{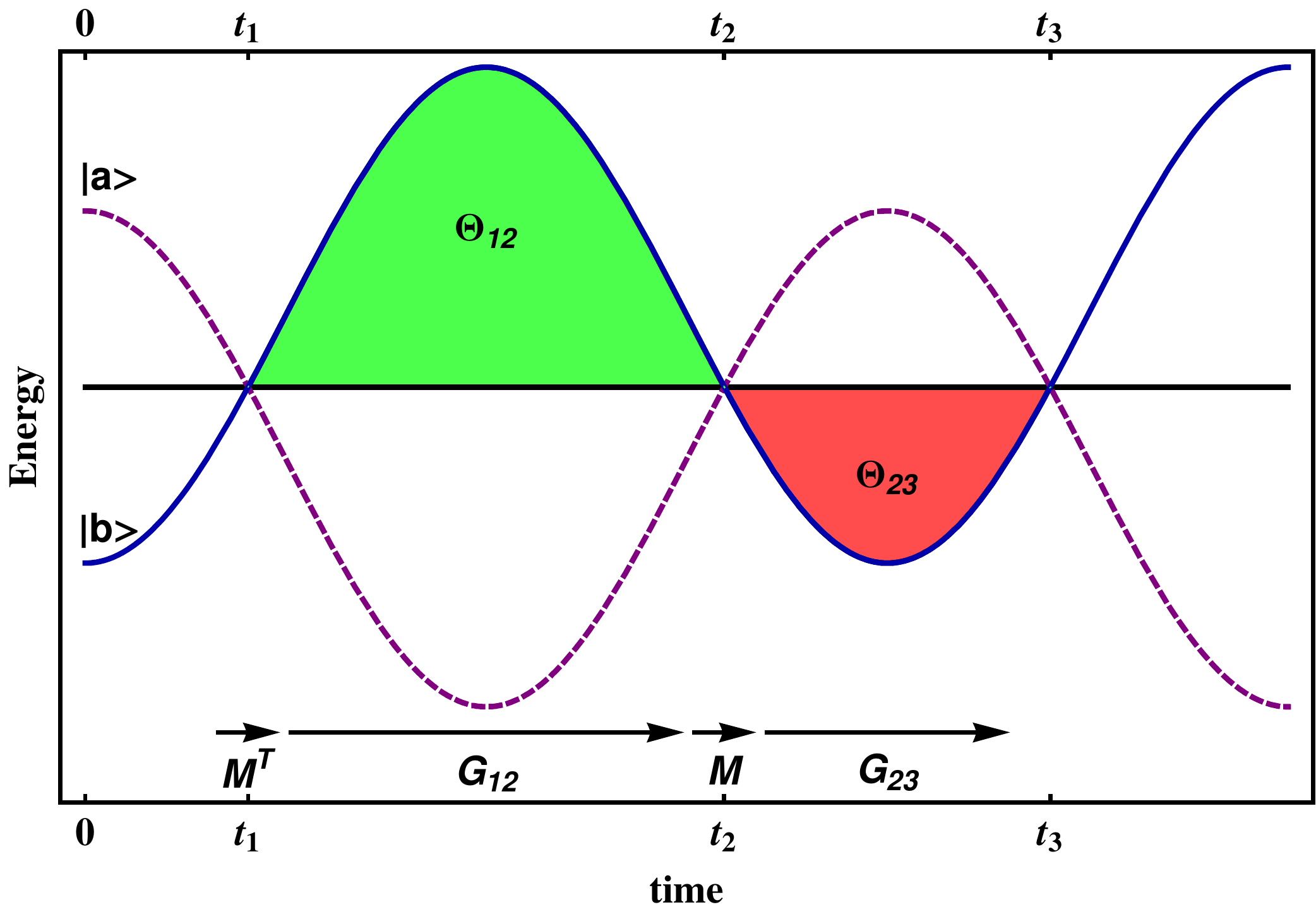}
\caption{(Color online) The energies of states $|a\rangle$ and $|b\rangle$ as a function of time during the RF-oscillation for the case of linear shifting Stark states. Indicated are the times where the two states cross, $t_1$, $t_2$ and $t_3$, and the matrices that describe the evolution of the wavefunction at the indicated time intervals: $\mathbf{M}^T$, $\mathbf{G}_{12}$,  $\mathbf{M}$ and $\mathbf{G}_{23}$. The green and red colored areas indicate the phase evolution of the state $|b\rangle$, $\Theta_{ij}$, which is contained in the $G_{ij}$ matrix.}\label{fig:oscstatesl}
\end{figure}

\begin{figure}[htb]
\includegraphics[width=0.42\textwidth]{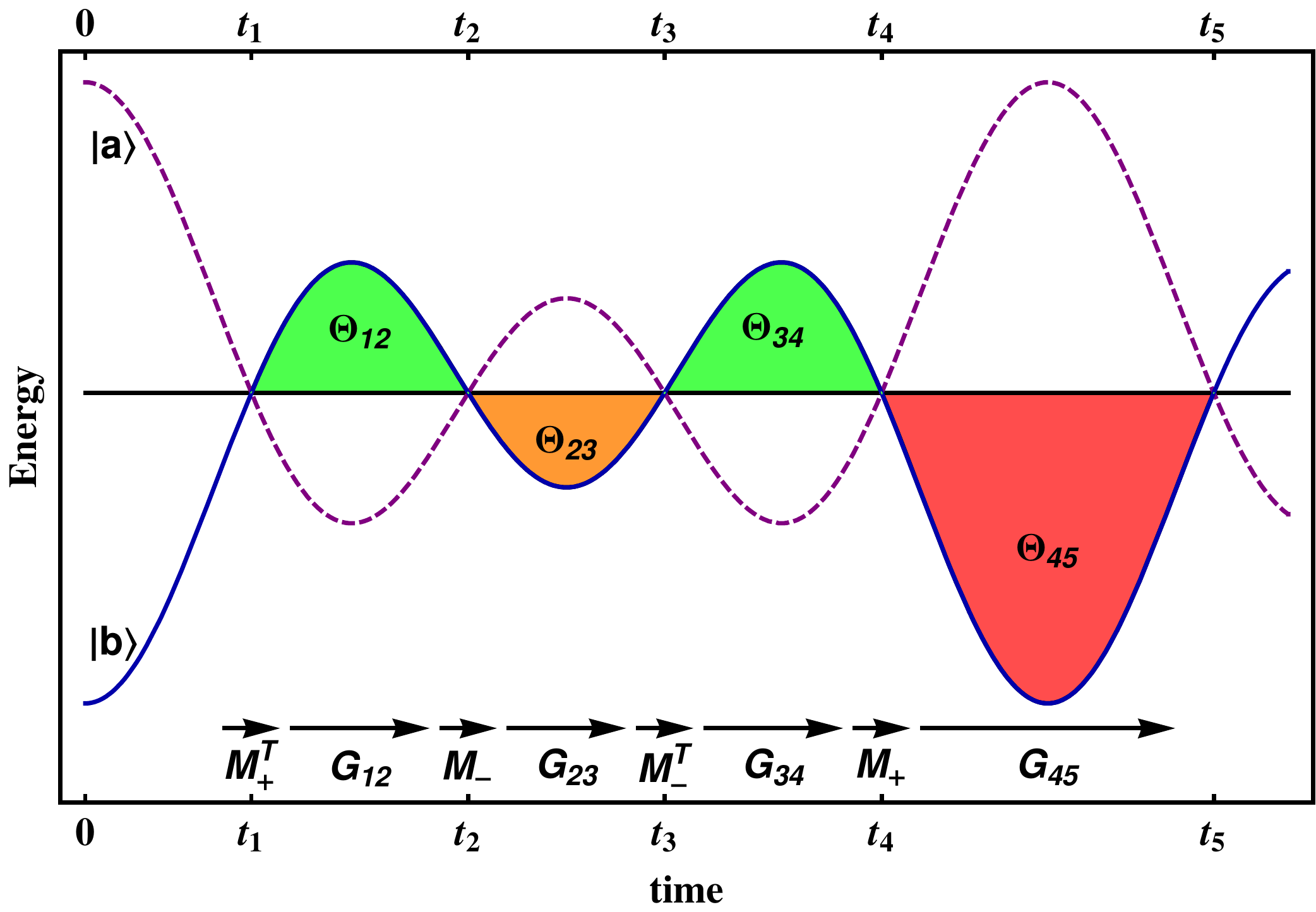}
\caption{(Color online) Similar to Fig.~\ref{fig:oscstatesl}; now for the case of quadratically shifting Stark states and large RF amplitudes, such that both the $+ F_{\mathrm{quad}}$ and the $- F_{\mathrm{quad}}$ crossing occur. At $t_1$ and $t_4$ the field is at the $+ F_{\mathrm{quad}}$ crossing and at $t_2$ and $t_3$ the field is at the $- F_{\mathrm{quad}}$ crossing. Furthermore, it is clearly visible that $\Theta_{12}$ and $\Theta_{34}$ are equal, they cover the section between $+ F_{\mathrm{quad}}$ and $- F_{\mathrm{quad}}$.}\label{fig:oscstatesq}
\end{figure}

To calculate the interference patterns we use the Landau-Zener-St\"{u}ckelberg (LZS) model \cite{landau,zener,stueckel}, as described in e.g.~\cite{Kayanuma1993} and apply it to the situation sketched in Fig.~\ref{fig:introstuck}. It is assumed that the avoided crossing (Eq.~\ref{eq:avcross}) is small compared to the range of occurring energies during an RF oscillation (Eq.\ \ref{eq:en_time_lin} or \ref{eq:en_time_quad}). In that case we can separate the acquired relative phase into a phase evolution due to energy differences and phase jumps due to avoided crossings. The Hamiltonian of the two-level system in a diabatic description is
\begin{equation}
\mathbf{H} =
\begin{pmatrix}
W_a(t) & V \\
V^\ast & W_b(t)
\end{pmatrix}\label{eq:hamsweep} ,
\end{equation}
where $W_a$ and $W_b$ are the unperturbed eigen-energies of respectively state $|a \rangle$ and $|b \rangle$, and V is the coupling between them.
In order to better exploit the symmetry of the problem, we choose a definition of energies that is slightly different from before: Here $W_b = - W_a = \frac{1}{2} W_2$, with $W_2$ defined in Eq.~\ref{eq:en_time_lin} or \ref{eq:en_time_quad}. The energies $W_a$ and $W_b$ during an RF oscillation are illustrated in Figs.~\ref{fig:oscstatesl} and \ref{fig:oscstatesq} for a specific choice of $F_S$ and $F_{RF}$. 

In the LZS model we describe the evolution of the wavefunction with transfer matrices. The transition at a level crossing is described by the transfer matrix \cite{Kayanuma1993}
\begin{equation}
\mathbf{M} =
\begin{pmatrix}
\sqrt{1\!-\!\epsilon} & \sqrt{\epsilon} \, e^{-i \phi}\\
-\sqrt{\epsilon} \, e^{i \phi} & \sqrt{1\!-\!\epsilon}
\end{pmatrix}\label{eq:transfer} .
\end{equation}
The matrix $\mathbf{M}$ is applicable for crossings where state $|a\rangle$ crosses $|b\rangle$ from the lower-energy side. In the opposite case the transpose of  $\mathbf{M}$, $\mathbf{M}^T$ should be used. The phase angle $\phi$ is the so-called Stokes phase \cite{ashhab}
\begin{equation}
\phi = \frac{\pi}{4} + \arg{\Gamma (1 - i \delta  )} + \delta (\ln ( \delta ) -1 )
\end{equation}
with values between 0 and $\pi /4$. And $\epsilon$ is the Landau-Zener adiabatic transfer probability 
\begin{equation}
\epsilon = 1- e^{-2 \pi \delta}.
\end{equation}
Both quantities depend on the adiabaticity parameter $\delta$
which in turn depends on the coupling strength $V$ and the rate of change of the difference of the diabatic energies
\begin{equation}
\delta = \frac{V^2}{| d(W_b - W_a )| / d t} .
\end{equation}
In the diabatic limit $\delta$ is small and subsequently $\epsilon$ is small as well, while $\phi$ approaches the value $\pi/4$. In the adiabatic limit $\epsilon$ approaches 1 and $\phi$ approaches 0. 

After the crossing, the fraction in state $|b \rangle$ develops a different phase compared to the fraction in state $|a \rangle$, before coming back to the crossing. Both phase evolutions are described by the matrix
\begin{equation}
\mathbf{G}_{ij} =
\begin{pmatrix}
e^{i \Theta_{i j}} & 0 \\
0 & e^{-i \Theta_{i j}}
\end{pmatrix},
\end{equation}
with $\Theta_{i j} \!=\! \int_{t_i}^{t_j} \sqrt{W_b(t)^2+V^2} d t \!\approx\! \int_{t_i}^{t_j} W_b(t) d t$. The approximation is valid for small $V$. Note that $\Theta_{i j}$ is a function of $F_S$ and $F_{RF}$. The matrices $\mathbf{M}$ and $\mathbf{G}$ which describe the phase development at and between the crossings are indicated in Figs.~\ref{fig:oscstatesl} and \ref{fig:oscstatesq}. The shaded surfaces indicate the phase of state $|b\rangle$ as it develops between the crossings. 

\begin{figure}[htb]
\includegraphics[width=0.48\textwidth]{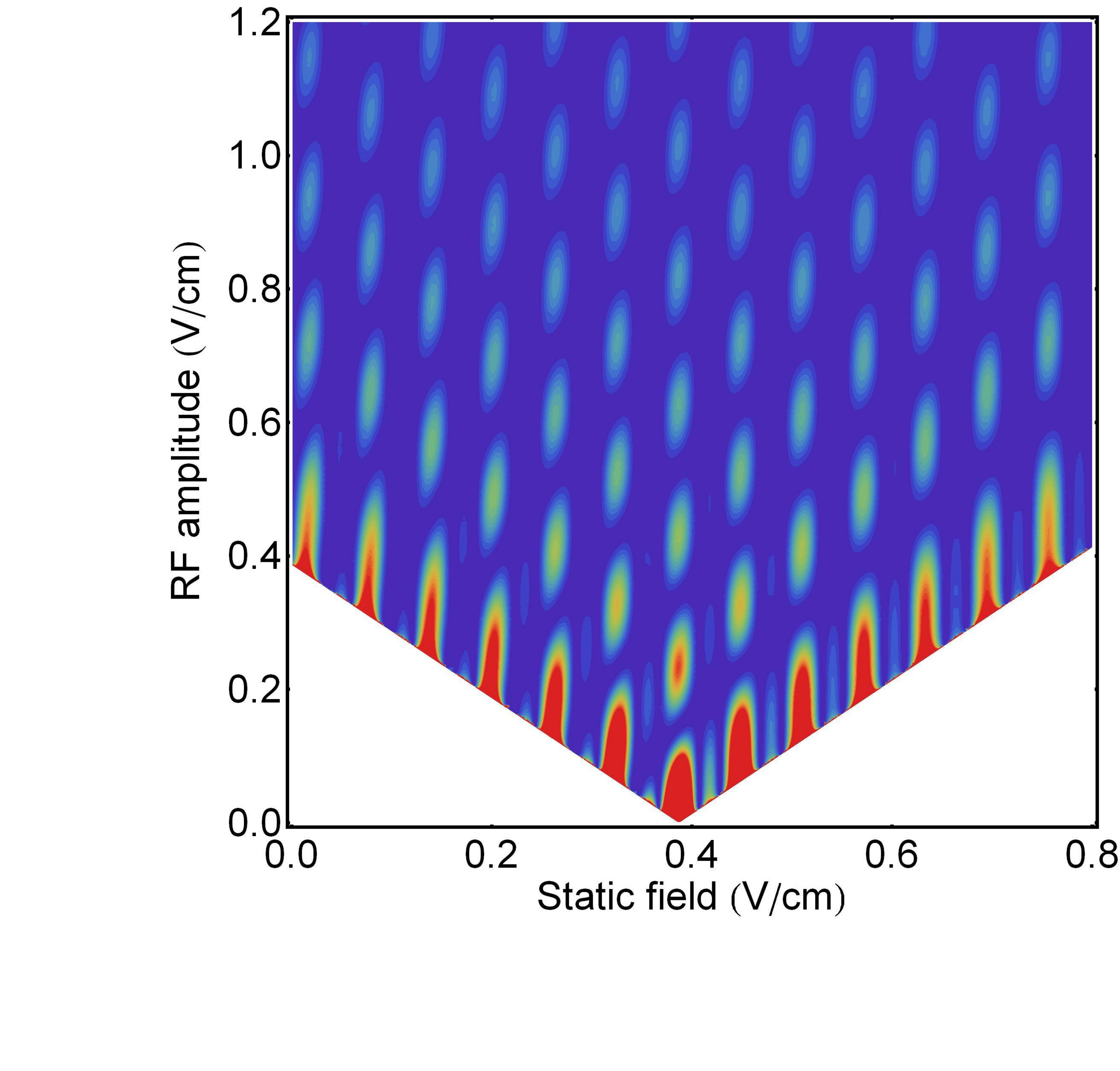}
\caption{(Color online) The population of state $|b \rangle$ as a function of the static field $F_S$ and the RF amplitude $F_{RF}$ (Eq.\ \ref{eq:bpoplin}) for $N=3$, where red is a large population and blue is a population of 0. }\label{fig:stucklin}
\end{figure}

For states with a linear Stark shift the situation is depicted in Fig.~\ref{fig:oscstatesl} and,  assuming that we start with all probability in $|a\rangle$, the amplitude of both states after $N$ oscillations is given by
\begin{equation}
(\mathbf{G}_{23} \:\! \mathbf{M} \:\! \mathbf{G}_{12} \:\! \mathbf{M}^T)^N  \left( \begin{array}{ccc} 1 \\ 0 \end{array} \right) = \mathbf{S}^N \left( \begin{array}{ccc} 1 \\ 0 \end{array} \right) ,
\end{equation}
where $\mathbf{S}$ describes the time evolution of the wavefunction over one RF oscillation.  The matrix $\mathbf{S}$ is unimodular ($\det \mathbf{S} \!=\! 1 $). Therefore we can write $\mathbf{S^N}$ in terms of Chebychev polynomials $U_n(\xi)$ \cite{Griffiths2001}
\begin{equation}
\mathbf{S}^N = \mathbf{S} \, U_{N\!-\!1}(\xi) - \mathbf{I} \, U_{N\!-\!2}(\xi) ,
\end{equation}
where $\xi= \mathrm{Tr}(\mathbf{S})$.
The solution of the full wavefunction in the linearly Stark-shifting case after $N$ oscillations can be expressed in terms of the above defined quantities. In view of the size of the expression, we just give the population of state $|b \rangle$
\begin{equation}
P_b = 4(1\!-\!\epsilon)\epsilon \sin^2(\Theta_{12}\!+\!\phi) \, U_{N\!-\!1}^2(\xi)
\end{equation}
with the trace
\begin{align}
\xi &= \cos(\Theta_{12}\!+\!\phi) \cos(\Theta_{23}\!-\!\phi) 
\nonumber \\
&- (1\!-\!2 \epsilon) \sin(\Theta_{12}\!+\!\phi) \sin(\Theta_{23} \!-\! \phi) .
\end{align}
In the near diabatic limit ($\epsilon \! \ll \! 1$) the population $P_b$ reduces to
\begin{equation}
P_b = 4 \epsilon \sin^2(\Theta_{12}\!+\!\phi) \, \frac{\sin^2(N(\Theta_{12}\!+\!\Theta_{23}))}{\sin^2(\Theta_{12}\!+\!\Theta_{23})} .\label{eq:bpoplin}
\end{equation}
In Fig.~\ref{fig:stucklin} we depicted the population of state $|b \rangle$ as a function of the static field $F_S$ and the RF amplitude $F_{RF}$ for $N=3$. Note the similarity with Fig.~\ref{fig:floqstucklin}. For larger $N$ the vertically elongated areas of large population become narrower.

\begin{figure}[htb]
\includegraphics[width=0.48\textwidth]{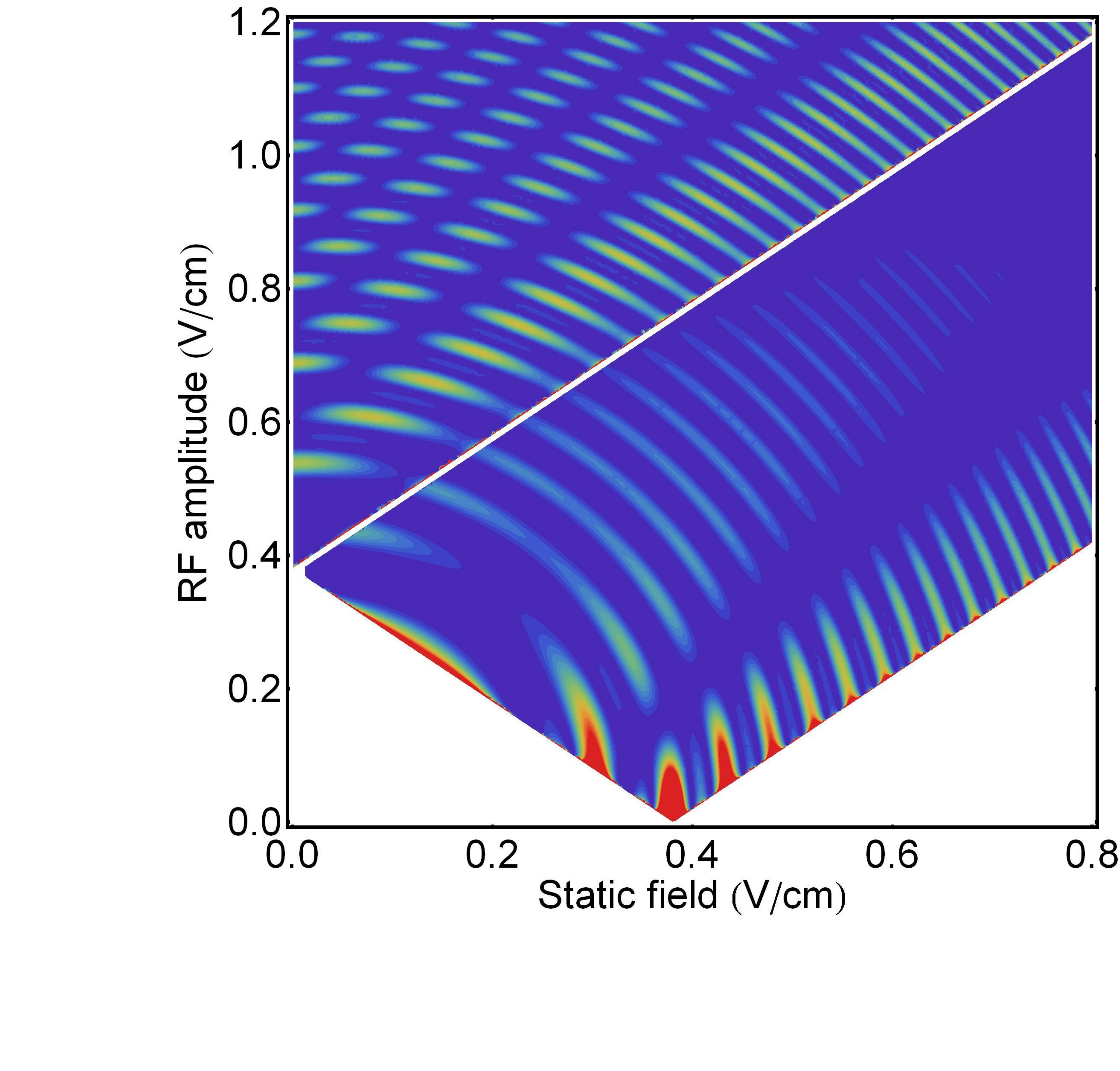}
\caption{(Color online) The population of state $|b \rangle$ as a function of the static field $F_S$ and the RF amplitude $F_{RF}$ (Eq.\ \ref{eq:bpopquad}) for $N=3$, where red is a large population and blue is a population of 0.}\label{fig:stuckquad}
\end{figure}

The quadratic case allows for a similar approach, but there is a twist. In cases where the states cross each other at only one field, either $+ F_{\mathrm{quad}}$ or $- F_{\mathrm{quad}}$ (Eq.~\ref{eq:fq}), the same equations are used as for the linear case. However, for large RF amplitudes it is possible that the two states cross at two different fields: $F \!=\! \pm F_{\mathrm{quad}}$. The speed at which each crossing is crossed is usually different, so we have different values for the adiabaticity parameter $\delta$ which will be labelled $+$ and $-$. The same holds for parameters that depend on $\delta$: $\phi$, $\epsilon$ and $\mathbf{M}$. The situation is illustrated in Fig.~\ref{fig:oscstatesq}. After N oscillations the wavefunction has evolved to
\begin{equation}
(\mathbf{G}_{45} \:\! \mathbf{M}_{_+} \:\! \mathbf{G}_{34} \:\! \mathbf{M}_{_-}^T \:\! \mathbf{G}_{23} \:\! \mathbf{M}_{_-} \:\! \mathbf{G}_{12} \:\! \mathbf{M}_{_+}^T)^N  \left( \begin{array}{ccc} 1 \\ 0 \end{array} \right) ,
\end{equation}
The parameter of the Chebychev polynomial now becomes
\begin{align}
\xi &= (1\!-\!\epsilon_{_-}) \cos(2 \Theta_{12} \!+\! \Theta_{23} \!+\!\Theta_{45}) 
\nonumber \\
&+ \epsilon_{_-} \cos(2\phi_{_-} \!+\!2\Theta_{12} \!-\! \Theta_{23}\!+\!\Theta_{45})
\nonumber \\
&+2\sin(\Theta_{45}\!-\!\phi_{_+}) \Big[ \epsilon_{_+} \sin(2\Theta_{12}\!+\!\Theta_{23}\!+\!\phi_{_+})
\nonumber \\
&-2\sin(\Theta_{23}\!-\!\phi_{_-}) \Big(\epsilon_{_+} \epsilon_{_-} \cos(2\Theta_{12}\!+\!\phi_{_-} \!+\!\phi_{_+})
\nonumber \\
&+ \sqrt{(1\!-\!\epsilon_{_-})\epsilon_{_-}(1\!-\!\epsilon_{_+})\epsilon_{_+}} \Big)\Big] ,
\end{align}
where we have used $\Theta_{12} \!=\! \Theta_{34}$. After $N$ oscillations the $|b \rangle$ population is
\begin{align}
P_b =& 4\Big[ (1\!-\!2\epsilon_{_+}) \sqrt{(1\!-\!\epsilon_{_-}) \epsilon_{_-}} \sin(\Theta_{23} \!-\! \phi_{_-})
\nonumber\\
&- \sqrt{(1\!-\!\epsilon_{_+})\epsilon_{_+}} \Big(\sin(2\Theta_{12}\!+\!\Theta_{23}\!+\!\phi_{_+})
\nonumber\\
-2\epsilon_{_-} \cos&(2\Theta_{12}\!+\!\phi_{_+}\!+\!\phi_{_-}) \sin(\Theta_{23}\!-\!\phi_{_-})\Big)\Big]^2 U_{N\!-\!1}^2(\xi)
\end{align}
For near-diabatic transitions $\epsilon_{_-},\epsilon_{_+} \! \ll \! 1$ the transferred population reduces to
\begin{align}
&P_b=4 \, \frac{\sin^2(N(2\Theta_{12}\!+\!\Theta_{23}\!+\!\Theta_{45}))}{\sin^2(2\Theta_{12}\!+\!\Theta_{23}\!+\!\Theta_{45})} 
\nonumber\\
&\Big[\sqrt{\epsilon_{_+}} \sin(2 \Theta_{12}\!+\!\Theta_{23}\!+\!\phi_{_+}) - \sqrt{\epsilon_{_-}} \sin(\Theta_{23}\!-\!\phi_{_-})\Big]^2 .\label{eq:bpopquad}
\end{align}
This $|b \rangle$ population is plotted as a function of the static field $F_S$ and the RF amplitude $F_{RF}$ for $N=3$ in Fig.~\ref{fig:stuckquad}. This figure shows a similar structure as Fig.~\ref{fig:floqstuckquad}.

In the expressions for the probability of state $|b \rangle$ (Eqs.\ \ref{eq:bpoplin} and \ref{eq:bpopquad}) we recognize two factors. Firstly, the familiar $\sin^2(N x)/\sin^2(x)$ factor which gives rise to the resonance condition $x = n \pi$; a very ubiquitous phenomenon. Working out this resonance condition we get exactly the same result as before: Eqs.~\ref{eq:reslin} and \ref{eq:resquad}. Secondly, we identify what we call the St\"uckelberg interference factor:
\begin{equation}\label{eq:stuck1}
I_1=4 \epsilon \sin^2(\Theta_{12}\!+\!\phi)
\end{equation}
for the linear case or the quadratic case where only one crossing occurs and 
\begin{equation}\label{eq:stuck2}
I_2=4 \big[ \sqrt{\epsilon_{_+}} \sin(2\Theta_{12}\!+\!\Theta_{23}\!+\!\phi_{_+})-\sqrt{\epsilon_{_-}}\sin(\Theta_{23}\!-\!\phi_{_-}) \big]^2
\end{equation} 
for two-crossings case. Note that these interference factors are independent of $N$ and that they occur already for one single oscillation.

In Fig.~\ref{fig:floqstucklin} we have depicted the regions of destructive interference. The green lines are the destructive interference regions for the situation described above. The red lines give the destructive interference regions for the case of a different starting point of the RF oscillation: between $t_1$ and $t_2$ (Fig.~\ref{fig:oscstatesl}). Note that in this case the phase difference that results in the interference (Eq. \ref{eq:stuck1}) involves $\Theta_{23}$ (the red area in Fig.\ \ref{fig:oscstatesl}) instead of $\Theta_{12}$ (the green area). This is because only the phase difference that has developed between the first two crossings is relevant; the last part gives just an overall phase factor to the wavefunction. 

In Fig.~\ref{fig:floqstuckquad} the double crossings case occurs in the upper left corner, above the upper blue dashed line. Below this blue dashed line (but above the lower two blue dashed lines) the one-crossing case is valid. In this lower part we have equivalently plotted the destructive interference in red and green. For the two-crossings case the interference factor $I_2$ (Eq. \ref{eq:stuck1}) involves the first three colored area's from Fig.~\ref{fig:oscstatesq}, $\Theta_{12}$, $\Theta_{23}$ and $\Theta_{34}$(which is equal to $\Theta_{12}$). So here the phase difference that has developed between $t_1$ and $t_4$ is relevant. The orange regions in Fig.~\ref{fig:floqstuckquad} correspond to the regions where this factor $I_2=0$. Starting between $t_2$ and $t_3$, the interference factor $I_2$ would contain $\Theta_{45}$ instead of $\Theta_{23}$ and the relevant phase difference has developed between $t_3$ and $t_6$ ($t_6$ is the moment of the crossing after $t_5$); this is indicated as the red regions in Fig.~\ref{fig:floqstuckquad}. Other starting points are possible, but not indicated in the figure since no new physics is involved.

The semi-classical Landau-Zener-St\"{u}ckelberg (LZS) model we described is an alternative to the Floquet model, described in sections \ref{sec:floquet} and \ref{sec:coupling}. Both models give approximately the same results, as we compare Fig.~\ref{fig:stucklin} with Fig.~\ref{fig:floqstucklin} and Fig.~\ref{fig:stuckquad} with Fig.~\ref{fig:floqstuckquad}. Secondly, the regions of destructive interference correspond nicely with the minima of the (generalized) Bessel function in Figs.~\ref{fig:floqstucklin} and \ref{fig:floqstuckquad}. This shows that both theories work well in the description of our measurements. The Floquet approach is valid for an infinite number of RF periods. In the experiment we work with an 8~MHz oscillation during 20~$\mu$s, i.e.\ 160 periods, so we believe that the Floquet approach is very well validated. The LZS description is not fully quantum mechanical - the exact time evolution near the crossings is not taken into account - and it does not give an answer outside the classically allowed regions, i.e.\ below the lower blue dashed lines in Figs.~\ref{fig:floqstucklin} and \ref{fig:floqstuckquad}. However, the LZS description might give more physical insight into the problem, such as the interpretation of the occurrence of maxima and minima in the interaction strength.


\section{Experiment}\label{sec:exp}

\subsection{Spectroscopy}\label{sec:spectr}

\begin{figure}[htb]
\includegraphics[width=0.45\textwidth]{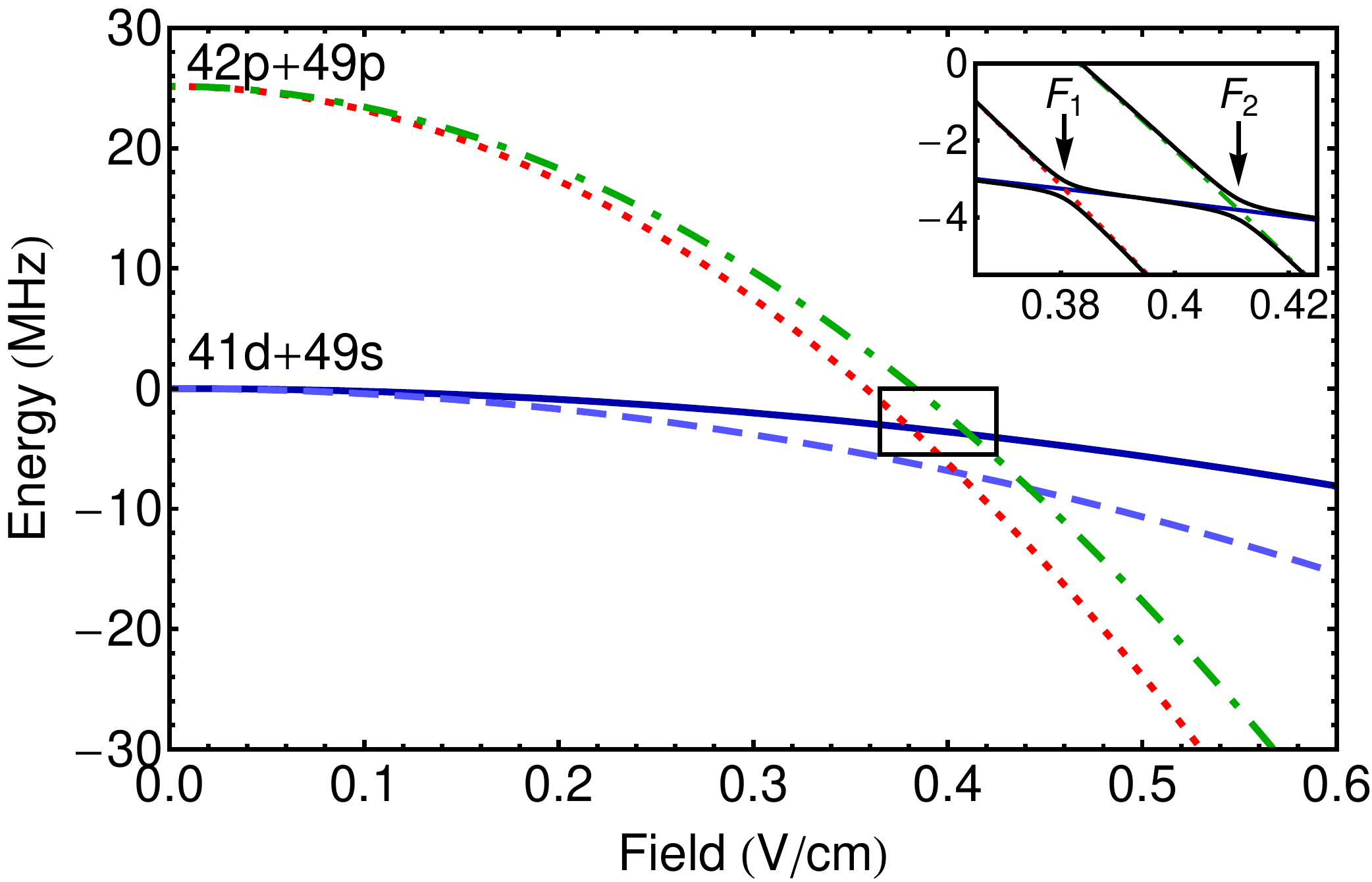}
\caption{(Color online) The two-atom energy levels of the 41d+49s and 42p+49p
system. In the inset we have zoomed in around the two relevant crossings, which become avoided crossings due to the dipole-dipole interaction.} \label{fig:enlev2atomagain}
\end{figure}

In the following part we will describe the performed experiments and compare the results with the theory presented above. For the two interacting states we have used two two-atom states, which are coupled through dipole-dipole interaction. We prepare atoms in the 41d state in a confined region of space, and some tens of microns away, we prepare atoms in the 49s state; in both cases by laser excitation. At a static field of approximately 0.4~V/cm the binding energy of one 41d$_{3/2}$ atom plus the binding energy of one 49s$_{1/2}$ atom equals the sum of binding energies of one 42p$_{1/2}$ atom and one 49p$_{3/2}$ atom. The two-atom energy level diagram is depicted in Fig.~\ref{fig:enlev2atomagain}. Both the 41d$_{3/2}$ and the 49p$_{3/2}$ atoms undergo Stark splitting of the two $|m_j|$ states, giving four two-atom states. The 42p$_{1/2,1/2}$ + 49p$_{3/2,1/2}$ state (notation: $n \ell_{j,|m_j|}$) is plotted as the dotted red line and the 42p$_{1/2,1/2}$ + 49p$_{3/2,3/2}$ state is plotted as the dash-dotted green line. The 41d$_{3/2,1/2}$ + 49s$_{1/2,1/2}$ state is shown as the solid blue line. The 41d$_{3/2,3/2}$ state is not excited by the laser, so the 41d$_{3/2,3/2}$ + 49s$_{1/2,1/2}$ state, depicted with a blue dashed line, does not play a role in our experiment. The energies and polarizabilities are calculated with the Numerov method \cite{PRA.20.2251} and we get $W_0=$25.15(13)~MHz. The transition
\begin{equation}
41\mathrm{d}_{3/2,1/2} + 49\mathrm{s}_{1/2,1/2} \leftrightarrow
42\mathrm{p}_{1/2,1/2} + 49\mathrm{p}_{3/2,1/2}\label{eq:reaction1}
\end{equation}
has a difference polarizability of $\alpha_1$ = 347.04(4) MHz/(V/cm)$^2$ and is therefore resonant at a field of $F_1$ = 0.3807(15) V/cm. The transition
\begin{equation}
41\mathrm{d}_{3/2,1/2} + 49\mathrm{s}_{1/2,1/2} \leftrightarrow
42\mathrm{p}_{1/2,1/2} + 49\mathrm{p}_{3/2,3/2}\label{eq:reaction2}
\end{equation}
has a difference polarizability of $\alpha_2$ = 297.40(4) MHz/(V/cm)$^2$ and is resonant at a field of $F_2$ = 0.4113(16) V/cm.

The states couple through resonant dipole-dipole interaction. The single-atom states have virtually no permanent dipole moment, but one atom oscillates due to the coherent 49s and 49p superposition and the other atom due to the 41d and 42p superposition. These superpositions of two states do have a dipole moment, a so called transition dipole moment $\boldsymbol{\mu}$, $\boldsymbol{\mu}_{sp} = \langle 49\mathrm{s} | \mathbf{r} | 49\mathrm{p} \rangle$ and $\boldsymbol{\mu}_{dp} = \langle 41\mathrm{d} | \mathbf{r} | 42\mathrm{p} \rangle$. Transition dipole moments oscillate with a frequency that is given by the energy difference of the contributing states. Both dipole moments are around 1000$a_0 e$ but can have various polarizations: $\boldsymbol{\mu}=\mu_z$ for $\Delta m_j = 0$ or $\boldsymbol{\mu}=\mu_x \pm i\mu_y$ for $\Delta m_j = \pm 1$. Note that, in contrast to the situation in a constant field,  here the sign of $m_j$ is relevant. The strength of the dipole-dipole interaction is given by the well-known expression
\begin{equation}
    V_\mathrm{dd} = \frac{\boldsymbol{\mu}_{sp} \cdot \boldsymbol{\mu}_{dp} - 3 (\boldsymbol{\mu}_{sp} \cdot \hat
    R) (\boldsymbol{\mu}_{sd} \cdot \hat R) }{R^3}\label{eq:potdipdip}
\end{equation}
and is approximately equal to $2\pi$~100~kHz for a distance of 25~$\mu$m. This results in typical values of the adiabaticity parameter $\delta$ between $10^{-5}$ and $10^{-3}$ (depending on the values of $F_S$ and $F_{RF}$). This means that the crossings are traversed in a highly diabatic manner. The adiabatic transfer probability $\epsilon$ is between $10^{-4}$ and $10^{-2}$, so the approximations eqs.\ \ref{eq:bpoplin} and \ref{eq:bpopquad} are well justified. The Stokes phase $\phi$ lies between 0.248$\pi$ and 0.25$\pi$. 

All 5 involved Rydberg states have a purely quadratic downwards Stark shift. This energy shift is due to the coupling with several $\ell$ states in a field. This means for the eigenstates in a field $\ell$ is not a good quantum number and the orbital angular momentum is not a conserved quantity. It is then not allowed to split the spatial and temporal dependence of the wavefunction (as in Eq.~\ref{eq:2_separate_wavefun}), because the field changes with time, and therefore $\psi(\textbf{r})$ changes with time. This is not a problem with linearly Stark shifted states, because here these states have $k$ as a good quantum number, the parabolic quantum number which determines the dipole moment of the state. So in the linear case $\psi(\textbf{r})$ does not change with the field (and therefore not with time). Fortunately, we need to use only very small fields. In the field of 0.4~V/cm we calculated the eigenstates of a whole range of Rydberg states and it appears that all involved states are more than 99.8\% pure, e.g.\ the population of the 49p part of the wavefunction we call 49p is 99.8\%. With this number we are confident that the separation of the space and time dependent part of the wavefunction is justified (see Eq.\ \ref{eq:wavefun2}).

In the experiment we explored the coupling strength as a function of the applied static field and the intensity of the radio frequency wave for this quadratically shifted system. In fact we made a few cuts though Fig.~\ref{fig:floqstuckquad}; a few horizontal scans, one scan along the vertical axis and finally several scans, where we followed a curved resonance line and could clearly observe the oscillations. The results (section \ref{sec:results}) are discussed after we have described the experimental setup in section \ref{sec:setup} and the simulations we have done on a system of multiple atom pairs in \ref{sec:sim}.


\subsection{Experimental Setup}\label{sec:setup}

To control the dipole-dipole interaction strength $V_\mathrm{dd}$ (Eq.~\ref{eq:potdipdip}) we confined the atoms in space by using focused laser beams for the creation of the Rydberg atoms, as described in \cite{Ditzhuijzen2008}. Two sets of pulsed dye-lasers excite ground-state atoms in a magneto-optical trap (MOT) to the Rydberg states. By using cold atoms, we ensure that the atoms hardly move on the timescale of the experiment. The cylinder-like volumes that contain the Rydberg atoms are approximately 15~$\mu$m in diameter, determined by the laser focus. The length is about 0.5~mm, determined by the diameter of the MOT cloud. Each volume contains about 25 interacting atoms. The distance between the two laser beams is chosen to be 25~$\mu$m. Several ms before the laser excitation the magnetic field for the MOT is switched off, because this field broadens the resonances due to position-dependent and $m_j$ dependent variation of the Zeeman effect.

The atoms are located between two field plates, on which the 8-MHz oscillating field is applied as well as an ionizing field ramp to detect the atoms and their states \cite{tombook}. The oscillating field is applied by an Agilent 33250A Arbitrary Waveform Generator, controlled by an NI Labview program. The distance vector between the cylinders is in the same direction as the electric field. We use the fraction of 49p atoms ($N_{49\mathrm{p}}/\left[N_{49\mathrm{s}}+N_{49\mathrm{p}}\right]$) as a measure for the interaction strength. The exact field is calibrated with a static field scan over the known resonances $F_1$ and $F_2$. The atoms are detected after 20~$\mu$s of interaction time. All data are averaged over 200 realizations. More details on the experimental setup can be found in \cite{Tauschinsky2008}, \cite{EPJD.40.13} and \cite{Ditzhuijzen2008}.


\subsection{Simulations for multiple atom pairs}\label{sec:sim}

In section \ref{sec:coupling} we derived expressions for the coupling strength as a function of the oscillating field. However, what we measure in the experiment is the 49p fraction, which is a measure for the 49p+42p state (pp-state) population. The population of this pp-state on resonance is given by $P_{pp} = \sin^2 \Omega t/2$,
since we start with 100\% in the sd-state. The interaction strength $\Omega$ is given by Eq.~\ref{eq:omega_n}, which contains the  interaction $V$ (Eq. \ref{eq:avcross}). This interaction is in this case the dipole-dipole interaction strength $V_\mathrm{dd}$ (Eq.\ \ref{eq:potdipdip}), which depends strongly on the atom-atom distance, which is restricted but not fixed. To match the theory better to the actual realistic situation, we performed simulations with multiple atom pairs. We have two elongated volumes, with the atoms sparsely distributed, giving varying distances for the atom pairs. $V$ is also angular dependent, this complicates the problem a lot, also because $\mu_{sp}$ and $\mu_{dp}$ have several possible polarizations. We will assume the angular dependence is averaged out and we use for $V_\mathrm{dd}$
\begin{equation}
V_\mathrm{dd} = \frac{\mu_{sd} \mu_{pp}}{\left(x^2+(y+d)^2+z^2\right)^{3/2}}\label{eq:vsimple2} ,
\end{equation}
with $d$ the distance between the two cylinders.

We randomly placed one particle in an elongated ellipsoid, Gaussian in 3D with full length 400~$\mu$m and radius 8~$\mu$m (both numbers are $1/\sqrt{e}$ widths) at 25~$\mu$m distance of a probe particle. For the interaction strength we used Eq.~\ref{eq:omega_n}, with $\Omega_0/2$ simply equal to $V_\mathrm{dd}$ (Eq.~\ref{eq:vsimple2}). For $\mu_{sd} \mu_{pp}$ we used $800^2 a_0^2 e^2 $. We varied the mixing angle, while keeping the effective field (Eq.~\ref{eq:feff}) constant on resonance for different $n$  and $\omega$=8~MHz. The mixing angle is defined as
\begin{equation}
\theta_F = \arctan \left(\frac{F_{RF}}{\sqrt{2} F_{S}} \right)\label{eq:mix} .
\end{equation}
In Fig.~\ref{fig:floqstuckquad} $\theta_F$ runs from 0 to $\pi/2$, where 0 corresponds to the horizontal axis and $\pi/2$ corresponds to the vertical axis. We evaluated the $|$pp$\rangle$ probability after 20~$\mu$s for 10000 random cases. The result is shown in Fig.~\ref{fig:angle} as the red dashed lines, where only the height is fitted to the experimental data.


\subsection{Results}\label{sec:results}

\begin{figure}[htb]
\includegraphics[width=.48\textwidth]{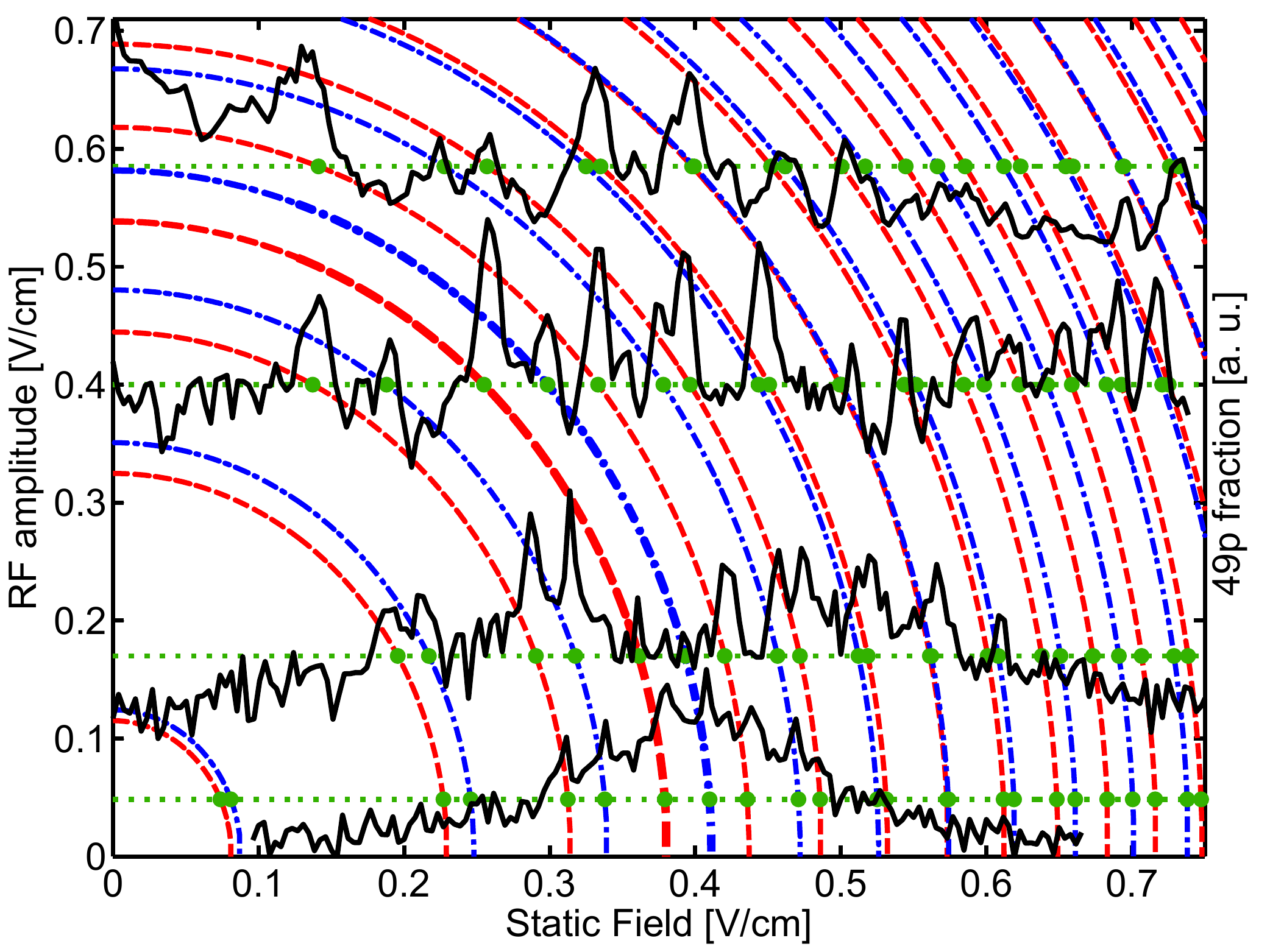}
\caption{(Color online) The 49p fraction in arbitrary units, depicted in black, is measured as a function of static field for different RF amplitudes. The red dashed and blue dash-dotted resonance lines correspond to respectively eq. \ref{eq:reaction1} and \ref{eq:reaction2} and run from the +3 photon transition (lower left corner) to the -14 or -11 photon transition (upper right corner). The 0-photon transition is depicted with thicker lines. The resonance peaks are expected where the green dotted lines cross with the red dashed and blue dash-dotted lines, indicated with green dots. }\label{fig:offset}
\end{figure}

As a first experiment we measured the 49p fraction, which is a measure for the dipole-dipole interaction strength, as a function of static field, for various values of the RF field, given in Fig.~\ref{fig:offset}. The measured data are depicted in black and are shown on top of a diagram of the resonances, like in Fig.~\ref{fig:floqstuckquad}. The red dashed lines belong to the transition in eq.\ \ref{eq:reaction1} and the blue dash-dotted lines belong to eq.\ \ref{eq:reaction2}. The green dotted lines depict the exact RF amplitudes of the performed scans, which are $F_{RF}$~=~0.049, 0.17, 0.4 and 0.585~V/cm. The green dots indicate where the resonances are expected, i.e.\ where the green dotted lines cross the red dashed and blue dash-dotted lines.

We observe that the positions of the peaks correspond nicely to the expected positions, the green dots. The exact field values $F_{RF}$ and $F_S$ have been slightly adjusted by a multiplication factor (maximum 2\% different from 1) to fit the expected positions, because the exact effective plate distance is not accurately known. The measurements show that indeed more sidebands occur for a stronger RF field, or in other words, more multi-photon transitions can be observed. With an amplitude of only 0.049~V/cm just two sidebands are populated, the one above and the one below the original states. We can also say that we have observed the $n \!=\! 1$ transition, where the two-atom system absorbs one RF photon, and we have the $n \!=\! 0$ transition, the original resonance, and there is the $n \!=\! -1$ transition, where the two-atom system emits one RF photon through stimulated emission. When we increase the RF amplitude to 0.17~V/cm we observe sidebands from $n \!=\! 2$ down to $n \!=\!- 5$. Surprisingly, at this field the original state is completely depleted: the $n \!=\! 0$ peaks are missing. At $F_{RF}$~=~0.4~V/cm the peaks around $n \!=\! -4$ are missing, while at $F_{RF}$~=~0.585~V/cm the peaks around $n \!=\! -9$ have disappeared. All these missing peaks are nicely predicted in Fig.~\ref{fig:floqstuckquad} by the green, red and orange areas or the white parts in the resonance curves. Here the generalized Bessel function and therefore the strength of the transition goes to zero, or in other words destructive St\"{u}ckelberg interference occurs.

\begin{figure}[htb]
\includegraphics[width=.48\textwidth]{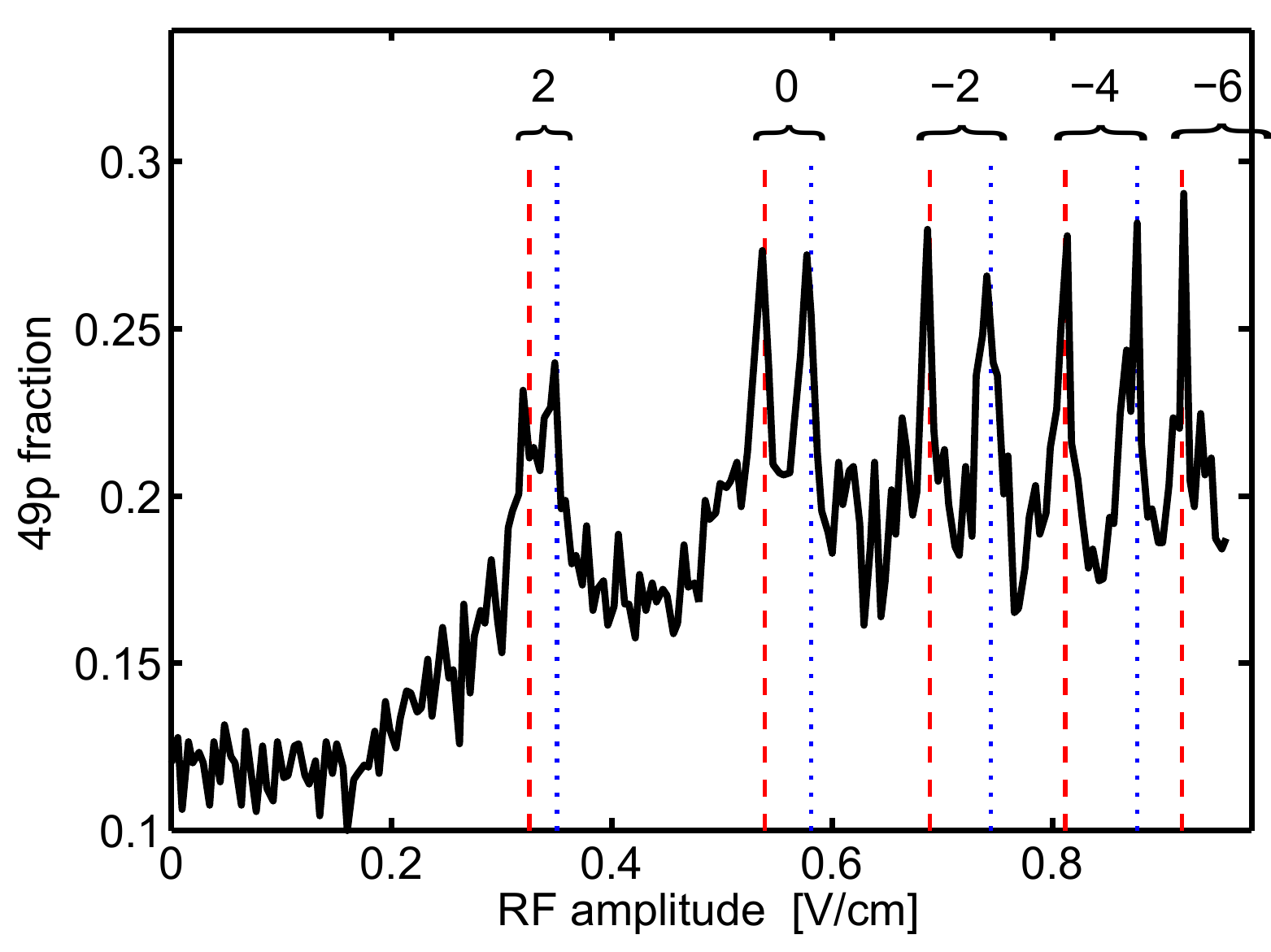}
\caption{(Color online) The 49p fraction as a function of RF amplitude at zero static field. The red dashed and blue dotted vertical lines depict the expected positions of the even photon transitions for both transitions given in Eq.\ \ref{eq:reaction1} and \ref{eq:reaction2}.}\label{fig:ampscan}
\end{figure}

As a second experiment, we measured the 49p fraction as a function of RF amplitude $F_{RF}$ for $F_S$=0. In other words, we made a vertical cut through Fig.~\ref{fig:floqstuckquad}, along the vertical axis. The results are depicted in Fig.~\ref{fig:ampscan}. The observed peaks fit the expected resonances, depicted as red dashed lines (transition in Eq.~\ref{eq:reaction1}) and blue dotted lines (transition in Eq.~\ref{eq:reaction2}) well; again a small adjustment of the field has been made. This figure shows in fact the pure AC-Stark shift of the states, which is $\frac{1}{4} \alpha F_{RF}^2$ (see Eq.~\ref{eq:en_sidebands_quad}). Interestingly, we observe only even-photon transitions. In the absence of a static field only an even number of photons can be transferred, due to the selection rules of photon transitions, as was the case in \cite{Tauschinsky2008} and is also visible in Fig.~\ref{fig:floqstuckquad}. Both atoms make a $\Delta \ell \!=\! \pm 1$ transition, and one photon -- with angular momentum 1 -- goes from one atom to the other. Adding an even number of photons the transition is possible, because then the total angular momentum of these RF photons can add up to zero. With an odd number of RF photons, however, some angular momentum remains, and the photons can not be absorbed by the two atoms. Already in a small static field odd-photon transitions are allowed, as can be seen in Fig.~\ref{fig:floqstuckquad} because the static field can contain some angular momentum (but no energy). In other words, with a static field other $\ell$ states are mixed in the atomic wavefunctions, as described in section \ref{sec:spectr}, and the photons can couple these parts of the wavefunction. Note that only a tiny fraction of other $\ell$ states is already enough to break up the selection rules. In other words, for a small field the radial part of the wavefunction is hardly changed, but the time dependent part is changed significantly.

\begin{figure}[htb]
\includegraphics[width=.48\textwidth]{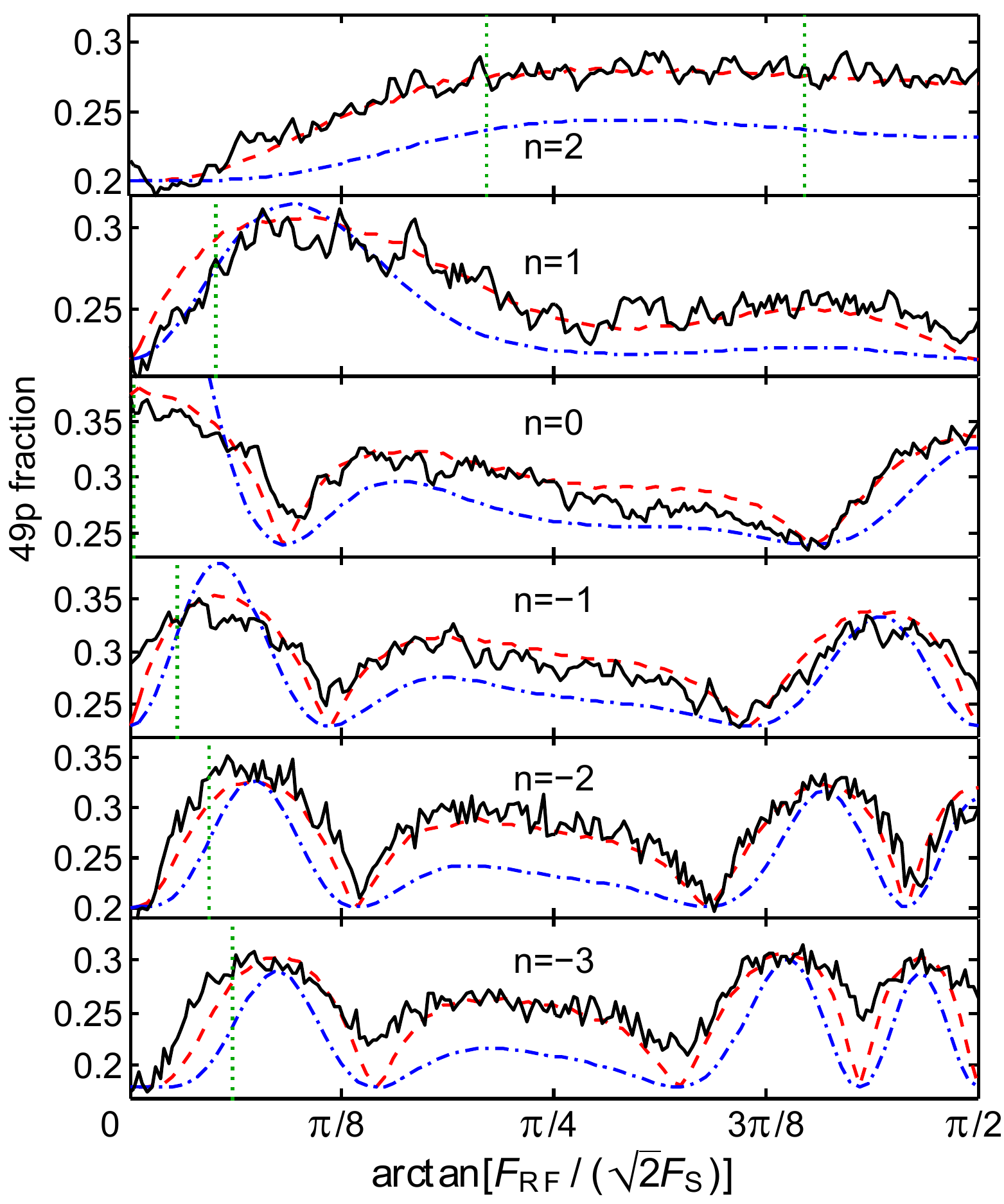}
\caption{(Color online) The 49p fraction measured for different mixing angles (Eq.\ \ref{eq:mix}). From left to right we have increasing $F_{RF}$ and decreasing $F_{S}$. The effective field is kept constant at the n-photon resonance for each plot. The red dashed lines show the result of the simulation discussed in section \ref{sec:sim}. The blue dash-dotted lines depict the squared generalized Bessel function (Eq.~\ref{eq:wavefunquad}) directly, which can be seen as a small transfer approximation. The green dotted vertical lines depict the boundaries of the classically allowed region and correspond to the lower blue dashed lines in Fig.~\ref{fig:floqstuckquad}.}\label{fig:angle}
\end{figure}

A third experiment illustrates the St\"{u}ckelberg oscillations most clearly. Here we keep the effective field (Eq.~\ref{eq:feff}) constant and fixed to a resonance and we vary the mixing angle (Eq.~\ref{eq:mix}). This requires adjustments of both the field amplitude and the static field. In Fig.~\ref{fig:angle} the result is depicted for mixing angles between 0 and $\pi/2$ for the $n \!=\! -3$ to $n\!=\!+2$ photon transitions of Eq.~\ref{eq:reaction1}. In fact the measurements run over the gray lines in Fig.~\ref{fig:floqstuckquad} starting on the bottom. The clear presence of the St\"{u}ckelberg interference patterns in the data, especially at the positions of the destructive interference, proves that our system is coherent for at least one period of the RF oscillation, i.e.\ 125~ns. Since the crossing is traversed mostly diabatically, many oscillations are needed to build up the pp population. It can not be proven from these data if this build up occurs coherently or incoherently. However, if we look at Figs.~\ref{fig:offset} and \ref{fig:ampscan}, we can estimate the ratio of peak width to peak distance to be approximately 5. This means that $N$ in Eq.~\ref{eq:bpopquad} is at least 5, so we can say that the interaction is coherent for at least 5 RF periods, i.e.\ 0.6~$\mu$s. In reality the coherence time could be longer, because the resonance peaks are probably broadened through other mechanisms.

The red dashed lines in Fig.~\ref{fig:angle} show the results of the simulations, described in section \ref{sec:sim}. Here only the height and the vertical offset are adjusted to fit the data. The simulations fit the data nicely, proving that this straightforward calculation reflects the real experimental situation quite well. The blue dash-dotted lines in Fig.~\ref{fig:angle} depict the squared generalized Bessel function directly (Eq.~\ref{eq:wavefunquad}). This reflects a situation of weak or short interaction, since the population of the pp-state then becomes $P_{pp} \!\approx\! V^2 J_n^2 t^2$. When the interaction is stronger, the pp probability oscillates between 0 and 1: the process is saturated. However, because multiple distances occur and therefore different interaction strengths, these oscillations are washed out. The result is that the peaks of the blue dash-dotted lines are broadened, as visible in the red dashed lines of the simulation. We see again that for small angles (DC-dominated cases) the population is fully in the $n\!=\!0$ state. With increasing RF amplitude the population in this state decreases and more and more higher order states are populated. For higher $|n|$ the start of the increasing population occurs at a larger RF amplitude (or larger mixing angle). This is also visible in the classical boundaries, depicted as green dotted lines, equivalent to the blue dashed lines in Fig.~\ref{fig:floqstuckquad}. Beyond these boundaries we observe the St\"{u}ckelberg oscillations. On the right hand side of the plot we see again that the odd photon transitions go to a minimum and the even photon transitions go to a maximum, because on this side the static field is zero and the selection rules apply as explained in connection with Fig.~\ref{fig:ampscan}.


\section{Conclusions}

We have studied both experimentally and theoretically interactions between atoms under the influence of an oscillating electromagnetic field. States with a quadratic Stark shift (polarizable states) show a fundamentally different behavior than states with a linear Stark shift (states with a permanent dipole moment). A comparison between the two cases has been made using the Floquet approach. Where in the linear case the resonance frequency depends purely on the static field and the coupling strength depends purely on the amplitude of the field, in the case of the quadratic shift, both the resonance frequency and the coupling strength depend on both the amplitude and the static offset of the field, and no separation of variables can be made. For the coupling strength in the quadratic case we invoked the generalized Bessel function. Its role is similar to the regular Bessel function for the linear case. The behavior of both functions is studied as a function of frequency and compared to the classical limit; the energy values that occur the most in the oscillating field correspond to the resonances with the strongest coupling strength, apart from some oscillatory behavior. The oscillatory behavior of the Bessel function as well as the generalized Bessel function can be explained in terms of St\"{u}ckelberg oscillations, which is an interference effect between the developed phases of the two interacting states.

In the experiment the resonance positions and the interaction strength fully fit the described theory. Destructive St\"{u}ckelberg interference is clearly observed, which proves that our system is coherent for at least one period of the RF oscillation, 125~ns. From the ratio of the mutual distance the peaks and the width of the peaks, the lower bound is increased by a factor of 5 to 0.6~$\mu$s. Simulations with multiple atom pairs at different distances, corresponding to our experimental situation of two cylinders of atoms, fit the observed interference patterns well.

\begin{acknowledgments}

We like to thank L.~D.~Noordam and R.~J.~C.~Spreeuw for stimulating discussions. This work is part of the research programme of the 'Stichting voor Fundamenteel Onderzoek der Materie (FOM)', which is financially supported by the 'Nederlandse Organisatie voor  Wetenschappelijk Onderzoek (NWO)'.

\end{acknowledgments}


\begin{thebibliography}{24}
\bibitem[{exp()}]{explain}
\bibinfo{note}{{The Bessel function $J_n(x)$ is roughly proportional to
  $Ai(2^{1/3} \frac{n-x}{n^{1/3}} ) $ for $n \ge x$. The Airy function $Ai(z)$
  drops of as $\exp (-\frac{2}{3} z^{3/2} )$ for large $z$. Combining this, we
  obtain that the larger-order Bessel function is smaller than some error
  $\epsilon$, with $\epsilon = \exp(-Q)$ if $n > x +Q^{2/3} x^{1/3}+\frac{1}{2}
  Q$. This drop-off limits the required summation range.}}
\end{thebibliography}
\end{document}